 \pdfoutput=1
\documentclass[12pt]{iopart}

\usepackage[pdftex]{graphicx}
\begin{document}
\def\A{{\cal A}}
\def\B{{\cal B}}
\title[Spinor states on almost $S^2$]{Spinor states 
on a curved infinite disc with non-zero spin-connection fields   }
\author{D  Lukman and N S Manko\v c Bor\v stnik}
\address{ Department of Physics, FMF, University of
Ljubljana, Jadranska 19, 1000 Ljubljana, Slovenia}
\ead{norma.mankoc@fmf.uni-lj.si}
 
\begin{abstract} 
In the paper of Lukman, Manko\v c Bor\v stnik, Nielsen (NJP 13 (2011) 103027) one step towards the 
realistic Kaluza-Klein[like] theories was made 
by presenting the case of a spinor in $d=(1+5)$ compactified on an (formally) infinite disc with 
the zweibein which makes a disc curved on an almost $S^2$ and with the spin connection 
field  which allows on such a sphere only one massless spinor state of a particular charge, 
coupling the spinor chirally to the corresponding Kaluza-Klein gauge field. The solutions for 
the massless spinor state were found for a range of spin connection fields, as well as the massive 
ones for a particular choice of the spin connection field. In this paper we present the 
massless and massive spinor states for the whole range of parameters of the spin connection
field which allow only one massless solution.
\end{abstract}



\maketitle


%
\section{Introduction}
\label{introduction}

This paper is the continuation of the paper entitled {\it "An effective two dimensionality" cases 
bring a new hope to the Kaluza-Klein[like] theories}~\cite{NHD}, in which the authors (the third one 
is H.B. Nielsen) proved on a toy 
model that when the space breaks from $M^{1 +(d-1)}$ to $M^{1+3}\times$  a non compact space 
there can exist in $M^{1+3}$ a massless solution for a spinor, chirally coupled to the 
corresponding Kaluza-Klein 
field, if the spin connection field and the vielbein on a non compact space are properly 
chosen~\footnote{
 T. Kaluza and O. Klein~\cite{kk} proposed that there is only the gravitational 
interaction and that gravity in $(1+4)$ dimensions manifests after the break of symmetries 
from $M^{1+4}$ to $M^{1+3}\times $ a circle as the ordinary gravity and the electromagnetic 
interaction. This very elegant idea  was later intensively studied by many 
authors~\cite{geogla,chofre,zee,salstr,mec,dufnilspop,daess,wet,zelenaknjiga,horpal} and extended 
to all the known interactions, in order  to prove that such a theory with the gravitational fields 
only -- that is with vielbeins and spin connections -- can manifest in the low energy region the 
so far observed particles and interactions. After E. Witten~\cite{witten} in his "no-go theorem" 
explained that the break of symmetries  from $M^{1+(d-1)}$ to $M^{1+3}\times $  a compact space 
causes that massless particles gain the mass of the scale of the break and can't therefore lead to 
the observed properties of particles and fields, the idea in its the most elegant version 
was almost abandoned. \\ 
One of us has been trying for long to develop the  {\it spin-charge-family} theory  
(N.S.M.B.)~\cite{norma92939495,holgernorma20023}, which would manifest in $d=(1+3)$ effectively as the 
{\it standard model} explaining  its assumptions: charges, families, gauge fields 
and scalar fields. In this theory spinors carry  in $d\ge 4$ nothing 
but two kinds of the spin (no charges) and interact with vielbeins and the two kinds of the 
spin connection fields. \\
The  {\it spin-charge-family} theory does accordingly 
share with the Kaluza-Klein[like] theories the problem of masslessness of the fermions 
before the electroweak break. The paper~\cite{NHD} as well as some of  the papers cited there,
is an attempt to prove that the Kaluza-Klein[like] theories might be the right way 
beyond the standard model in non compact spaces.}~\footnote{The authors Manko\v c and Nielsen~\cite{hnkk06}
achieved masslessness of spinors on an infinite disc with appropriate boundary conditions.}. 
Although we did just assume the two fields used (not showed how they are generated, say, 
due to the presence of some spinor fields in interaction with some vielbeins and spin connections not 
mentioned in the paper) yet we succeeded  to find the analytical solutions for a  massless spinor for the 
whole allowed interval of the parameter of the (chosen) spin connection field and the spectrum 
for massive solutions for a particular choice of the parameter. 

In this paper we present the masses and states for spinors living on $M^{1+3}\times$  an almost $S^2$ 
sphere on which the parameter of the chosen spin connection field is allowed to change in the interval, which 
assures normalisable solutions. We prove that the presented solutions are the only normalisable ones 
and that they form a complete basis.

The reader, who is not willing to follow the introduction into the equations of motion for the two 
unknown functions $\mathcal{A}_{n}$ and $\mathcal{B}_{n+1}$, which determine the spinor states 
on an almost $S^2$ sphere with a particular spin connection on it, can jump 
directly to the coupled  equations of motion~(\ref{main1}, \ref{maineps}) of the first order and 
further to 
the two differential equations of motion of the second order for the unknown functions 
$\mathcal{A}_{n}$ (\ref{mainsecA}) and $\mathcal{B}_{n+1}$ (\ref{mainsecB}), the 
solutions of which are interesting by themselves. (She or he might like to see which physical problem 
initiated the equations, at the end, if at all.) 
%
First of the two equations~(\ref{mainsecA}) namely  simplifies, when putting the parameter 
($\varepsilon$) of the spin connection field equal to $0$, to the equation for the Legendre
polynomials, while  solutions for the second equation~(\ref{mainsecB}) are expressible, when using  
the solutions of the first one~(\ref{mainsecA}) and one  of the two coupled first order differential 
equations of motion (\ref{main1}), in quite an elegant way.  For a general choice of the parameter 
($\varepsilon$) from the allowed interval, determining the spin 
connection field,  the solutions are expressible with the finite sum of the associate Legendre polynomials.

We shall here briefly repeat 
the part  from the paper~\cite{NHD} needed to come to~(\ref{main1}). 

Let us start with the action for massless (Weyl) spinors~\cite{NHD} living on the manifold $M^{1+5}$ 
\begin{eqnarray}
\label{action}
S&=& \int d^{d}x \,  {\cal L}_{W}\,, \nonumber\\
{\cal L}_{W} &=& \frac{1}{2} [(\psi^{\dagger} E \gamma^0 \gamma^a p_{0a} \psi) + 
(\psi^{\dagger} E \gamma^0\gamma^a p_{0 a}
\psi)^{\dagger}]\,,\nonumber\\
p_{0a }&=& f^{\alpha}{}_a p_{0\alpha} +  \frac{1}{2E}\, \{ p_{\alpha}, E f^{\alpha}{}_a\}_- \,,
\nonumber\\  
         p_{0\alpha}&=& p_{\alpha} - 
                     \frac{1}{2}\, S^{ab}  \, \omega_{ab\alpha}\,.                     
\end{eqnarray}
Here $f^{\alpha}{\!}_{a}$ are the vielbeins~\footnote{$f^{\alpha}{}_{a}$ are inverted 
vielbeins to 
$e^{a}{}_{\alpha}$ with the properties $e^a{}_{\alpha} f^{\alpha}{\!}_b = \delta^a{}_b,\; 
e^a{}_{\alpha} f^{\beta}{}_a = \delta^{\beta}_{\alpha} $. 
Latin indices  
$a,b,..,m,n,..,s,t,..$ denote a tangent space (a flat index),
while Greek indices $\alpha, \beta,..,\mu, \nu,.. \sigma,\tau ..$ denote an Einstein 
index (a curved index). Letters  from the beginning of both the alphabets
indicate a general index ($a,b,c,..$   and $\alpha, \beta, \gamma,.. $ ), 
from the middle of both the alphabets   
the observed dimensions $0,1,2,3$ ($m,n,..$ and $\mu,\nu,..$), indices from 
the bottom of the alphabets
indicate the compactified dimensions ($s,t,..$ and $\sigma,\tau,..$). 
We assume the signature $\eta^{ab} =
diag\{1,-1,-1,\ldots,-1\}$.
} (the gauge fields of the infinitesimal generators of translation) and  $\omega_{ab\alpha}$ 
the spin connections (the gauge fields of $S^{ab}= \frac{i}{4}(\gamma^a \gamma^b - 
\gamma^b \gamma^a)$). 
The notation $\{ A, B\}_{\mp}$ means $\{AB \mp BA \}$. 
Correspondingly the Lagrange density for a spinor reads
\begin{eqnarray}
{\cal L}_{W}&=& \psi^{\dagger}\, \gamma^0 \gamma^a E \{f^{\alpha}{}_a p_{\alpha} +
\frac{1}{2E} \{p_{\alpha},f^{\alpha}{}_a E\}_-   -\frac{1}{2} S^{cd}  \omega_{cda} 
 \}\psi.
\label{weylL}
\end{eqnarray}
Space $M^{1+5}$ has after the break the symmetry of $M^{1+3} \times $ an infinite disc with the zweibein
on the disc 
\begin{eqnarray}
e^{s}{}_{\sigma} = f^{-1}
\pmatrix{1  & 0 \cr
 0 & 1 \cr},
f^{\sigma}{}_{s} = f
\pmatrix{1 & 0 \cr
0 & 1 \cr}\,,
\label{fzwei}
\end{eqnarray}
where  
\begin{eqnarray}
\label{f}
f &=& 1+ (\frac{\rho}{2 \rho_0})^2 = \frac{2}{1+\cos \vartheta}=  \frac{2}{1+ x}\,,
\nonumber\\ 
x^{(5)} &=& \rho \,\cos \phi,\quad  x^{(6)} = \rho \,\sin \phi, \quad E= f^{-2}\,,\nonumber\\
\nonumber\\
0 \le &\phi& \le 2 \pi\,,\quad 0\le \rho \le \infty\,,-1 \le \cos \theta \le 1\,, \nonumber\\
&& \quad x=\frac{1-(\frac{\rho}{2 \rho_0})^2}{1+(\frac{\rho}{2 \rho_0})^2}\,,
 \quad \quad -1 \le x \le 1\,.
\end{eqnarray}
We use indices $s,t=(5,6),$ to describe the flat index in the space of an infinite plane, and 
$\sigma, \tau = (5), (6), $ to describe the Einstein index.  
With $\phi$ we denote the angle of rotations around  the axis perpendicular to the disc.
The angle $\vartheta$ is the ordinary azimuth  angle on a sphere. 
The last relation follows  from $ds^2= 
e_{s \sigma}e^{s}{}_{\tau} dx^{\sigma} dx^{\tau}= f^{-2}(d\rho^{2} + \rho^2 d\phi^{2})$.
Here $\rho_0$ is the radius of an almost $S^2$ sphere.
The volume of this non compact sphere is finite, equal to $V= \pi\, (2 \rho_0)^2$.  The symmetry 
of $S^2$ is a symmetry of $U(1)$ group. 

There is also a spin connection field on a disc, chosen to be
\begin{eqnarray}
  f^{\sigma}{}_{s'}\, \omega_{st \sigma} &=& i F\, f \, \varepsilon_{st}\; 
  \frac{e_{s' \sigma} x^{\sigma}}{(\rho_0)^2}\, , \quad 
 0 <2F \le 1\, 
  ,\quad s=5,6,\,\,\; \sigma=(5),(6)\,. 
\label{omegas}
\end{eqnarray}
It compensates in the particular case when $2F=1$  the term 
$\frac{1}{2E} \{p_{\sigma},f^{\sigma}{}_s\}_{-} $ for   spinors  of
one of the handedness~(\ref{main1}) 
and determines the  spectrum of a massless and massive states to be $(m\rho_0)^2
= l(l+1)$, with $l=\{0,1,2,\cdots\}$. 


We require  normalizability (needed like in any of this kind of 
problems to lead to operative solutions) of states $\psi$ on the disc 
\begin{eqnarray}
\label{normalizability}
\int_{0}^{2\pi}\,d \phi \; \int_{0}^{\infty}\;\rho d\rho \psi^{\dagger} \psi < \infty. 
\end{eqnarray}
We assume to have no gravity in $d=(1+3)$ ($f^{\mu}{}_m = \delta^{\mu}_m$ and  
$\omega_{mn\mu}=0$ for $ m,n=0,1,2,3,\ldots; \;,  \mu =0,1,2,3, \ldots $). 

It is proven in~\cite{NHD} 
that there are chiral fermions on this almost sphere with the 
spin connection field on it without including any extra fundamental gauge fields: A massless spinor  
of only one handedness, and correspondingly mass protected, which couples to the corresponding Kaluza-Klein 
fields.

The equations of motion for spinors (the Weyl equations), which follow from the Lagrange 
density~(\ref{weylL}) when assuming that there is no gravity in $d=(1+3)$,  are then
\begin{eqnarray}
&&\{E\gamma^0 \gamma^m p_m + E f \gamma^0 \gamma^s \delta^{\sigma}_s  ( p_{0\sigma} 
+  \frac{1}{2 E f}
\{p_{\sigma}, E f\}_- )\}\psi=0,\quad {\rm with} \nonumber\\
&& p_{0\sigma} = p_{\sigma}- \frac{1}{2} S^{st}\omega_{st \sigma}\,,
\label{weylE}
\end{eqnarray}
with $f$ from~(\ref{f})
and with $ \omega_{st \sigma}$ from~(\ref{omegas}).
Let us add that the vielbein curling the disc into an almost $S^2$ does not break the rotational 
symmetry on the disc, it breaks the translation  symmetry 
after making a choice of the northern and southern pole.

The  solution  of the equations of motion (\ref{weylE}) for a spinor   
in $(1+5)$-dimensional space, which breaks into  
$M^{(1+3)}$ and a non compact $S^2$, is a superposition
of all  four ($=2^{6/2 -1}$) states of a single Weyl representation. (We kindly ask the 
reader to see the technical details  about how to write a Weyl representation 
in terms of the Clifford algebra objects after making a choice of the Cartan subalgebra,  
for which we take $S^{03}, S^{12}$ and $S^{56}$, in~\cite{NHD,holgernorma20023}.)
In our technique 
one spinor representation---the four 
states, which all are the eigen states of the chosen Cartan subalgebra with the eigen 
values~$\frac{k}{2}$---are  
the following four products of projectors  $\stackrel{ab}{[k]}$ and nilpotents~$\stackrel{ab}{(k)}$: 
\begin{eqnarray}
\varphi^{1}_{1} &=& \stackrel{56}{(+)} \stackrel{03}{(+i)} \stackrel{12}{(+)}\psi_0\,,\nonumber\\
\varphi^{1}_{2} &=&\stackrel{56}{(+)}  \stackrel{03}{[-i]} \stackrel{12}{[-]}\psi_0\,,\nonumber\\
\varphi^{2}_{1} &=&\stackrel{56}{[-]}  \stackrel{03}{[-i]} \stackrel{12}{(+)}\psi_0\,,\nonumber\\
\varphi^{2}_{2} &=&\stackrel{56}{[-]} \stackrel{03}{(+i)} \stackrel{12}{[-]}\psi_0\,,
\label{weylrep}
\end{eqnarray}
where  $\psi_0$ is a vacuum state for the spinor state.
%
%
The most general wave function  
$\psi^{(6)}$ obeying~(\ref{weylE}) in $d=(1+5)$ can now be written as
\begin{eqnarray}
\psi^{(6)} = \A \,{\stackrel{56}{(+)}}\,\psi^{(4)}_{(+)} + 
\B \,{\stackrel{56}{[-]}}\, \psi^{(4)}_{(-)}\,. 
\label{psi6}
\end{eqnarray}
$\A$ and $\B$ depend on $x^{\sigma}$ and $\psi^{(4)}_{(+)}$ 
and $\psi^{(4)}_{(-)}$  determine the spin 
and the coordinate dependent parts of the wave function $\psi^{(6)}$ in $d=(1+3)$ 
\begin{eqnarray}
\psi^{(4)}_{(+)} &=& \alpha_+ \; {\stackrel{03}{(+i)}}\, {\stackrel{12}{(+)}} + 
\beta_+ \; {\stackrel{03}{[-i]}}\, {\stackrel{12}{[-]}}\,, \nonumber\\ 
\psi^{(4)}_{(-)}&=& \alpha_- \; {\stackrel{03}{[-i]}}\, {\stackrel{12}{(+)}} + 
\beta_- \; {\stackrel{03}{(+i)}}\, {\stackrel{12}{[-]}}\,. 
\label{psi4}
\end{eqnarray}
If one uses $\psi^{(6)}$ of~(\ref{psi6}) in~(\ref{weylE}), separates dynamics in 
$(1+3)$ and on $S^2$, expresses $x^{(5)}$ and $x^{(6)}$ from~(\ref{f}) and takes 
the zweibein from~(\ref{fzwei}, \ref{f}) and the spin connection from~(\ref{omegas}), 
the equation~(\ref{weylE}) transforms as follows 
%
%
%
\begin{eqnarray}
\label{weylErho}
\fl
if \, \{ e^{i \phi 2S^{56}}\, [\frac{\partial}{\partial \rho} + \frac{i\, 2 S^{56}}{\rho} \, 
(\frac{\partial}{\partial \phi}) -  \frac{1}{2 \,f} \, \frac{\partial f}{\partial \rho }\, 
(1- 2F \, 2S^{56})\,]
\, \} \, \psi^{(6)}
+ \gamma^0 \gamma^5 \, m \, \psi^{(6)}=0\,.
\end{eqnarray}
%
$\psi^{(6)}$ can be chosen to be  the eigen function of the total angular momentum
$M^{56}= 
-i \frac{\partial}{\partial \phi} + S^{56}$
\begin{eqnarray}
\psi^{(6)}= {\cal N}_{n}\, ({\cal A}_{n}\, \stackrel{56}{(+)}\, \psi^{(4)}_{(+)}  
+ {\cal B}_{n+1}\, e^{i \phi}\, \stackrel{56}{[-]}\, \psi^{(4)}_{(-)})\, e^{in \phi}\,,
\label{mabpsi}
\end{eqnarray}
with the property
\begin{eqnarray}
M^{56}\psi^{(6)}= (n+\frac{1}{2})\,\psi^{(6)}\,.
\label{mabx}
\end{eqnarray}
Here ${\cal N}_{n}$ is the normalization constant. 
Taking into account that $S^{56} \stackrel{56}{(+)}= \frac{1}{2} \stackrel{56}{(+)}$, and 
$S^{56} \stackrel{56}{[-]}= -\frac{1}{2} \stackrel{56}{[-]}$,  and
introducing the coordinate $x$ 
from~(\ref{f}) instead of $\rho$, we 
end up with the equations of motion
 for ${\cal A}_n$ and ${\cal B}_{n+1}$ as follows  
\begin{eqnarray}
\label{main1}
&&-i \left[ -\sqrt{1-x^2}\left(\frac{d}{d x}
      -\frac{n+1}{1-x^2}\right)
    -\frac{1+2F}{2}\sqrt{\frac{1-x}{1+x}}\right]\mathcal{B}_{n+1}
+m\rho_0\mathcal{A}_n =0\, , \nonumber\\
&&-i \left[ -\sqrt{1-x^2}\left( \frac{d}{d x}
      +\frac{n}{1-x^2}\right)
    -\frac{1-2F}{2}\sqrt{\frac{1-x}{1+x}}\right]\mathcal{A}_{n}
+m\rho_0\mathcal{B}_{n+1} =0\,,\nonumber\\
&& -1 \le x \le 1 \,, \quad 0\le (1-2F)< 1\,.
\end{eqnarray}
The interval of the coordinate $x$, $(-1 \le x \le 1)$, corresponds to  the interval 
of the coordinate $\rho$, $(0 \le \rho \le \infty )$.  
The spinor part ($\stackrel{56}{(+)}$, $\stackrel{56}{[-]}$) and the angular part $e^{in \phi}$, 
both manifest orthogonality.

For any $F$ within the interval $0 <2F \le 1$ only one normalisable massless spinor state on $S^2$ 
exists~\cite{NHD}. In the  particular case that $2F=1$ the spin connection term 
$- S^{56} \omega_{56 \sigma} $ compensates the term  $\frac{1}{2Ef} \{ p_{\sigma},E f \}_-$ for the 
left handed spinor with respect to $d=1+3$, while for the spinor of the opposite handedness, 
again with respect to $d=1+3$,  the spin connection term doubles the term 
$\frac{1}{2Ef} \{p_{\sigma},E f\}_-$:  The term $\sqrt{\frac{1-x}{1+x}}$  
in~(\ref{main1})  is multiplied by $1$  in the first equation and 
by $0$ in the second equation.   

While the massless solution was found in~\cite{NHD} for the whole interval of the parameter 
$F$, $(0\le (1-2F)< 1)$, the massive spectrum was presented  only for the particular 
case  $2F=1$. 

In this paper we find the spectrum  for the whole interval of $F$, $(0\le (1-2F)< 1)$.

\section{Solutions of the equations of motion  for spinors}
\label{equations}

We look in this section for the spectrum of the equation of motion (\ref{main1}) for an arbitrary choice 
of the parameter $F$ within the interval $0 <2F \le 1$. 
We allow only normalisable solutions~(\ref{psi6}, \ref{mabpsi}). Taking into account that $E \rho d\rho =dx$ it follows for the normalizability 
requirement 
\begin{eqnarray}
2\pi \, \int^{1}_{-1} \,dx\, \A^{\star}_{n} \A_{n} && < \infty\,, \nonumber\\
2\pi \, \int^{1}_{-1} \,dx\, \B^{\star}_{n} \B_{n} && < \infty\,. 
\label{normalcono}
\end{eqnarray}

Let us, for simplicity, introduce a new parameter $2F = 1-2 \varepsilon $ and 
rewrite~(\ref{main1}) with this new parameter $\varepsilon$ 
\begin{eqnarray}
\label{maineps}
&&-i \left[ -\sqrt{1-x^2}\left( \frac{d}{d x}
      -\frac{n+1}{1-x^2}\right)
    -(1-\varepsilon)\sqrt{\frac{1-x}{1+x}}\right]\mathcal{B}_{n+1}
+m\rho_0\mathcal{A}_n =0,\nonumber\\
&&-i \left[ -\sqrt{1-x^2}\left( \frac{d}{d x}
      +\frac{n}{1-x^2}\right)
    -\varepsilon\sqrt{\frac{1-x}{1+x}}\right]\mathcal{A}_{n}
+m\rho_0\, \mathcal{B}_{n+1} =0,\,\nonumber\\
&& 0 \le \varepsilon <\frac{1}{2}\,.
\end{eqnarray}
One easily finds the massless ($m\,\rho_0 = 0$) normalisable solutions~\cite{NHD} for 
$\mathcal{A}_{n}$ for any 
$0 \le \varepsilon <\frac{1}{2}$ ($0<2F\le 1$)
\begin{equation}\label{Am0}
\mathcal{A}_n = \mathcal{A}_o\, \rho^n f^\varepsilon = \mathcal{A}_o\,
(1-x)^{\frac{n}{2}}(1+x)^{-\frac{n}{2}-\varepsilon}\,\rho_{0}^n \, 2^{n+\varepsilon}.
\end{equation}
The corresponding massless solution for $\mathcal{B}_{n}$ 
\begin{equation}\label{Bm0}
\mathcal{B}_{n}= \mathcal{B}_o\, \rho^{-n} f^{1-\varepsilon} 
= \mathcal{B}_o \, (1-x)^{-\frac{n}{2}}(1+x)^{\frac{n}{2}+\varepsilon-1}\, 2^{-n +1-\varepsilon} 
\, \rho_{0}^{-n}
\end{equation}
is not normalisable. 
The normalisation condition namely requires~\cite{NHD}
\begin{eqnarray}
&&{\rm for}\; \A_{n}: \, -1 < n < (1- 2 \varepsilon)\,, \nonumber\\
&&{\rm for}\; \B_{n}: (1- 2 \varepsilon) < n < 1, \quad n \;\; {\rm is \;\; an \;\;integer}\,.
\label{masslesseqsolf1}
\end{eqnarray}
For $(0 \le \varepsilon <\frac{1}{2})$ 
one immediately sees that this is for an integer $n$  
possible only for $n=0$, $\A_{n}= \A_{0}$~(\ref{Am0}) and $\B_{n}=0$~(\ref{Bm0}). 

\Eref{masslesseqsolf1} tells us that the strength $F$ ($= (\frac{1}{2}- \varepsilon$)) 
of the spin connection field $\omega_{56 \sigma}$  makes a choice between the two massless 
solutions $\A_0$ and $\B_0$. 

To look for  massive solutions of~(\ref{maineps})  we express the eigen states in terms 
of the associate Legendre polynomials $P^{l}_{n}$, which solve the equation
   \begin{eqnarray}
   \label{pln}
   \{(1-x^2) \frac{d^2}{dx^2}-2x \frac{d}{dx}  + (l(l+1)-
   \frac{n^2}{1-x^2})  \} P^{l}_{n}=0\,.
   \end{eqnarray}
 One can prove that the Legendre polynomials form a normalisable 
 \begin{eqnarray}
 \label{Plnnorm}
 \int_{-1}^1 \,P^{l}_n \,P^{l'}_n \,dx &=&\frac{2}{2 l+1}\,
      \frac{(l+n)!}{(l-n)!}\, \delta_{l l'}\,, \nonumber\\
 \int_{-1}^1 \,P^{l}_n\, P^{l}_{n'} \,\frac{dx}{1-x^2}&=& \frac{1}{n}\,
      \frac{(l+n)!}{(l-n)!} \,\delta_{nn'}\,, \nonumber\\
l, l' \ge n\,; \quad n,n'\ge 0\,,
\end{eqnarray}
 and complete set~\cite{WG,CS,NHD}  in the interval $-1 \le x\le 1$ only for an  integer $l$ 
 ($P^{\lambda}_{\nu}$  is not square integrable for $\lambda$ not an integer $l$) and an 
 integer $n$ ($P^{\lambda}_{\nu}$  is not square integrable for $\nu$ not an integer $n$). 
 Due to the orthogonality of the spinor and angular parts by themselves in~(\ref{mabpsi}) 
 we need only the orthogonality relation of the first line in~(\ref{Plnnorm}).

Let us transform~(\ref{maineps}), which are the two coupled first order equations, into the two 
second order equations for $\mathcal{A}_n $ and $\mathcal{B}_{n+1}$ 
\begin{eqnarray} 
\label{mainsecA}
\fl
(1 - x^2) \frac{d^2}{dx^2}\mathcal{A}_n -2 x \frac{d}{dx}\mathcal{A}_n +(m\rho_0)^2 \mathcal{A}_n\nonumber\\
-\left\{\frac{n^2}{1-x^2} +\frac{1}{1+x} \varepsilon (1 + x +2 n)  + 
  \varepsilon^2 \frac{1-x}{1+x} \right\}\mathcal{A}_n =0\,,
\end{eqnarray}
\begin{eqnarray}
\label{mainsecB}
\fl
(1 - x^2) \frac{d^2}{dx^2}\mathcal{B}_{n+1} - 2 x
\frac{d}{dx}\mathcal{B}_{n+1} +(m\rho_0)^2 \mathcal{B}_{n+1}\nonumber\\
 +\left\{-\frac{ (n+1)^2}{1 - x^2}  
+\frac{2n(1-\varepsilon)}{1+x} 
+\varepsilon(1-\varepsilon)\frac{1-x}{1+x} 
\right\} \mathcal{B}_{n+1}=0\,.
\end{eqnarray}

We immediately see that for $\varepsilon =0$ the normalisable solutions of~(\ref{mainsecA}) are 
the associate Legendre polynomials ${\cal A}^{l}_{n} = P^{l}_{n}$  of~(\ref{pln}) 
and that the mass spectrum must be discrete, $(m \rho_0)^2= l (l+1)$,  ($l=\{0,1,2,\cdots\}$), 
in order that the solutions fulfil the normalisability condition  of~(\ref{normalcono}).   
The solutions ${\cal B}^{l}_{n}$ of~(\ref{mainsecB})  
follow from the second equation of (\ref{maineps})  once ${\cal A}^{l}_{n}$ are known.

To find the normalisable solutions of~(\ref{mainsecA}, \ref{mainsecB})  for any $\varepsilon$ 
in the interval $0 \le \varepsilon <\frac{1}{2}\,$ 
one could expand solutions in terms of $P^{l}_{n}$, ${\cal A}_{n}= \sum_{l= n, \infty} \, \alpha^{l}_{n} 
\,P^{l}_{n} $  and ${\cal B}_{n}= \sum_{l= n, \infty} \, \beta^{l}_{n} 
\,P^{l}_{n} $. The recurrence  relations for the coefficients $ \alpha^{l}_{n} $  and 
$\beta^{l}_{n} $, which follow when using these two expansion in~(\ref{mainsecA}, \ref{mainsecB}) 
and taking into account~(\ref{pln}), are presented 
in~\ref{usefulequations} (equations~(\ref{alpha}, \ref{beta})). It is, however, very 
difficult to see when using these two expansions in terms of the Legendre polynomials for which values of 
the mass term $m \rho_0$ are solutions normalisable. 

It is much more convenient to find a normalisable and useful ansatz by evaluating the behaviour of 
solutions of~(\ref{mainsecA}, \ref{mainsecB}) 
at $x \to 1$  and at $x \to - 1$. 
This is done in subsect.~\ref{yz}. We find for ${\cal A}_{n}$  
and $0\le \varepsilon < \frac{1}{2}$  
\begin{eqnarray}
\label{Ananzatzinfty}
\mathcal{A}_n &=& (1+x)^{-\varepsilon}\, \sum_{l=n}^\infty \, a^{l}_n\, P^{l}_n\,.
\end{eqnarray}
For ${\cal B}_{n+1}$ and $0\le \varepsilon < \frac{1}{2}$ we find
\begin{eqnarray}
\label{Bnanzatzinfty}
\mathcal{B}_{n+1} &=& (1+x)^{-\varepsilon+ \frac{1}{2}}\, \sum_{l=n}^\infty \, b^{l}_n\, P^{l}_n\,.
\end{eqnarray}
We shall make use of~(\ref{Ananzatzinfty}) and use the second equation of~(\ref{maineps}) to find the solutions 
for $\mathcal{B}_{n+1}$.

Using this ansatz in~(\ref{mainsecA}), and taking into account that the first line  
in~(\ref{mainsecA}) is for each $l$ equal to ($-l(l+1) + (m \rho_0)^2$), the recursion relation 
for the coefficients $a^{l}_n$ follows
\begin{eqnarray}
\label{aln}
\fl a^{l+1}_{n}  \frac{l+n+1}{2 l+3}\,\left[
  (m\rho_0)^2-(l+2)(l+1+2 \varepsilon)\right]= \nonumber\\
-a^{l}_n \,\left[ (m\rho_0)^2-l(l+1)-2 n \varepsilon\right]\nonumber\\
-a^{l-1}_n \frac{l-n}{2l-1} \left[(m\rho_0)^2-(l-1)(l-2\varepsilon)\right].
\end{eqnarray}
Let us have a look at how does this recursive relation manifest at $l\to \infty$ for  a fixed $n$ and a fixed 
$(m \rho_0)^2$. One finds
\begin{eqnarray}
\label{alninfty}
a^{l+1}_{n} \, \frac{1}{2} &=& -a^{l}_n -a^{l-1}_n \, \frac{1}{2}\,,\quad {\rm or}\nonumber\\
 a^{l}_n &=& -  \frac{a^{l+1}_n + a^{l-1}_n}{2}\,.
\end{eqnarray}
This means that  coefficients $a^{l}_n$ are for  large $l \to \infty$ up to a sign all equal.
Since the Legendre polynomials $P^{l}_{n}$ are for each $n$ orthogonal, while their normalisation 
factor~(\ref{Plnnorm}) is proportional to $\frac{2}{2l+1}$, 
and 
\begin{eqnarray}
\label{alninftypln}
&& \sum_{l=n}^{\infty}\, \frac{2}{2 l+1} \to \infty\,, 
\end{eqnarray}
this means that for the normalisable solution~(\ref{Ananzatzinfty}) only a finite sum over 
Legendre polynomials is allowed.
This further means that there must be  $l_0$, $l_0\ge n$,  which closes the sum. 
Let us accordingly require for each $l \ge l_{0}$ that 
\begin{eqnarray}
\label{alnfinite}
&& a^{l_{0}+2}_{n} = 0 = a^{l_{0}+1}_{n}\,, \quad a^{l_{0}}_{n}\ne 0\,.  
\end{eqnarray}
It follows  from~(\ref{aln})
\begin{eqnarray}
\label{alnmass}
&& 0 = 0 -a^{l_0}_n \,\frac{l_0 +1 -n}{2l_0 +1} \,[(m\rho_0)^2-l_0(l_0 +1 - 2\varepsilon)]\,,
\end{eqnarray}
which only can be fulfilled  for 
\begin{eqnarray}
\label{mass}
&& (m\rho_0)^2=l_0(l_0 +1 - 2\varepsilon)\,, \nonumber\\
&& 0\le \varepsilon < \frac{1}{2}\,.
\end{eqnarray}
\Eref{mass} determines the mass spectrum of spinors on the infinite disc curved into an almost $S^2$ 
and with the spin connection from~(\ref{omegas}). 
The recursion relation in~(\ref{aln}) determines the solution $\mathcal{A}^{l_0}_n$ for a particular $l_0$ 
\begin{eqnarray}
\label{Ananzatzl0}
\mathcal{A}^{l_0}_n &=& (1+x)^{-\varepsilon}\, \sum_{l=n}^{l_0} \, a^{l}_n\, P^{l}_n\,.
\end{eqnarray}
The corresponding $\mathcal{B}_{n+1}$ can be found if 
using the second of equations in~(\ref{maineps}) and the relations among Legendre polynomials, 
presented in~\ref{legendre}  as 
\begin{eqnarray}
\label{Bnanzatzl0}
\fl
\mathcal{B}^{l_0}_{n+1} = -\frac{i\,(1+x)^{-\varepsilon}}{\sqrt{l_0(l_0 +1-2\varepsilon)}
\sqrt{1-x^2)}}\, \times \nonumber\\
 \sum_{l=n}^{l_0} \, a^{l}_n\,\{ n\, P^{l}_n + \frac{1}{(2 l+1)}[(l+n)(l+1)P^{l-1}_{n}-
l(l-n+1)P^{l+1}_{n}]\}\,.
\end{eqnarray}
Solutions~\Eref{mabpsi} can now be written as 
\begin{eqnarray}
&&\psi^{(6) l_0}_{n}= {\cal N}^{l_0}_{n}\, \left({\cal A}^{l_0}_{n}\, \stackrel{56}{(+)}\, \psi^{(4)}_{(+)}  
+ {\cal B}^{l_0}_{n+1}\, e^{i \phi}\, \stackrel{56}{[-]}\, \psi^{(4)}_{(-)}\right)\, e^{in \phi}\,,\nonumber\\
&&(m \rho_0)^2= l_0 (l_0 +1 -2\,\varepsilon)\,,\nonumber\\
&& 0\le \varepsilon < \frac{1}{2}\,,
\label{mabpsil0}
\end{eqnarray}
with ${\cal A}^{l_0}_{n}$ from~(\ref{Ananzatzl0}, \ref{aln}) and ${\cal B}^{l_0}_{n+1}$ from~(\ref{Bnanzatzl0}).
Since the associate Legendre polynomials form a complete normalisable set, as we pointed out when 
discussing the properties of solutions of~(\ref{pln}, \ref{Plnnorm}), the solutions presented 
in~(\ref{mabpsil0}) form the only normalisable solutions of~(\ref{maineps}, \ref{mainsecA}
).

For a special case of $\varepsilon =0$ we reproduce the spectrum presented in the ref.~\cite{NHD} with 
$(m \rho_{0})^2 = l_0(l_0 +1)$. The recursive relation~(\ref{aln}) allows in this case  only one 
nonzero coefficient, namely $a^{l_0}_{n}$, which we immediately see if putting $a^{l_0+1}_{n}=0$ for 
$\varepsilon=0$ and obtain $a^{l_0-1}_{n}=0.$ Consequently all the rest of
coefficients are  equal to zero due to~(\ref{aln}). 
The solution  for $\varepsilon=0$ then reads
\begin{eqnarray}
\fl\psi^{(6) l_0}_{n+1/2} = 
{\cal N}^{l_0}_{n+1/2} \, \biggl(
\stackrel{56}{(+)} \psi^{(4)}_{(+)} \nonumber\\ + 
\frac{i}{2 \sqrt{l_0(l_0+1)}} \, \stackrel{56}{[-]} \, \psi^{(4)}_{(-)} \, e^{i \phi} \, \sqrt{1-x^2}\,
(\frac{d}{dx} \, -\frac{n}{1-x^2})\, \biggr)
 \cdot \,e^{i n \phi} \, P^{l_0}_n \,.
\label{knsol}
\end{eqnarray}
For $l_0=0$ the (only) massless solution $\psi^{(6) 0}_{1/2}=$ ${\cal N}^{0}_{1/2} \, 
\stackrel{56}{(+)} \psi^{(4)}_{(+)}$ follows. 

In sect.~\ref{conclusion} the solutions~(\ref{mabpsil0}) are presented 
and their properties discussed  for several choices of $\varepsilon$ in the interval 
$0\le \varepsilon < \frac{1}{2}$.

\subsection{Behaviour of solutions of equations of motion at $x=-1$ and $x=1$}
\label{yz}

In this section we study behaviour of solutions of~(\ref{mainsecA}, \ref{mainsecB})
in the vicinity of the two ends of the interval  $-1 \le x \le 1$. We find a normalisable ansatz 
for solutions  of~(\ref{maineps}) by evaluating the contributions of $\int\,dx |{\cal A}_n|^2$ 
and $\int\,dx \,|{\cal B}_n|^2$ in the vicinity of both ends of the interval. 

Let us start with  $x\to 1$ and let us expand ${\cal A}_n$ as 
\begin{eqnarray}
\label{abto1}
{\cal A}_{n}   (x \to 1) &\sim& \,(1 + \frac{1}{2}(\varepsilon -(m \rho_0)^2)(1-x) + \cdots)\,,\nonumber\\
{\cal B}_{n+1} (x \to 1) &\sim& \,\frac{-i\sqrt{2}}{m \rho_0}\, (1-x)^{\frac{1}{2}}\,(\frac{1}{2} (m \rho_0)^2 
 \cdots)\,.
\end{eqnarray}
The coefficient in the expansion in $(1-x) $ was  found by checking
the validity of~(\ref{mainsecA})  
and~(\ref{mainsecB}) when $(1-x) \to 0 $. 
One easily checks that the contribution of these ansatzes from the very vicinity of $x= 1$, 
for any small $\eta $, is finite
\begin{eqnarray}
\label{abto1int}
&&\int_{0}^{\eta}\, d(1-x) |{\cal A}_{n}|^2 < \infty \,, \nonumber\\
&&\int_{0}^{\eta}\, d(1-x) |{\cal B}_{n+1}|^2 < \infty \,. 
\end{eqnarray}
Similarly we proceed  at $x \to -1$ with ansatzes, the coefficients of which were found 
by using these ansatzes in~(\ref{mainsecA}, \ref{mainsecB}), 
\begin{eqnarray}
\label{abto-1}
{\cal A}_{n} (x \to -1)  &\sim&    \, (1+x)^{- \varepsilon}\, (1 - 
\frac{(m \rho_0)^2}{2(1-2\varepsilon)}\, (1+ x) +  \cdots)\,,\nonumber\\
{\cal B}_{n+1} (x \to -1) &\sim&   \,i\sqrt{2}\,(1+x)^{-\varepsilon + \frac{1}{2}}
\, \frac{m \rho_0}{2(1-2\varepsilon)} (1+  \cdots)\,. 
\end{eqnarray}
The contribution of the integrals below in the vicinity of $x=-1$ are finite only for 
$0\le \varepsilon < \frac{1}{2}$
\begin{eqnarray}
\label{abto1int1}
&&\int_{0}^{\eta}\, d(x+1) |{\cal A}_{n}|^2 < \infty \,, \nonumber\\
&&\int_{0}^{\eta}\, d(x+1) |{\cal B}_{n+1}|^2 < \infty \,,\nonumber\\
&& 0\le \varepsilon < \frac{1}{2}\,.
\end{eqnarray}
Accordingly the ansatz of~(\ref{Ananzatzl0}) 
follows 
\begin{eqnarray}
\label{Ananzatzl0sub}
\mathcal{A}^{l_0}_n &=& (1+x)^{-\varepsilon}\, \sum_{l=n}^{l_0} \, a^{l}_n\, P^{l}_n\,,\quad
\mathcal{B}_{n+1} = (1+x)^{-\varepsilon+ \frac{1}{2}}\, \sum_{l=n}^\infty \, b^{l}_n\, P^{l}_n\,.
\end{eqnarray}

\section{Discussions and conclusions}
\label{conclusion}

We present in this paper the spectrum of a spinor living on a manifold $M^{1+5}$, which breaks 
to $M^{1+3}\times$  an infinite disc with the zweibein~(\ref{fzwei}) 
which curls the infinite disc into an almost $S^2$ and with the spin connection field on a 
disc~(\ref{omegas}) which in the whole interval of the parameter $\varepsilon$~(\ref{maineps}) 
allows only one massless spinor state.  Accordingly there  exists in $M^{1+3}$  after the break 
a massless solution for a spinor which is chirally coupled to the corresponding Kaluza-Klein field.

We find the whole mass spectrum~(\ref{mass}) of  
solutions and the corresponding normalisable~(\ref{normalcono}) spinor 
states~(\ref{mabpsil0}, \ref{Ananzatzl0}, \ref{Bnanzatzl0}), 
which form a complete set of states.

The coupled first order differential equations of motion~(\ref{maineps}) for the two functions 
${\cal A}_n$ and ${\cal B}_{n+1}$,  which determine solutions~(\ref{psi6}, \ref{mabpsil0}), 
lead to  two second order differential equations~(\ref{mainsecA}, \ref{mainsecB}), 
 one of which is for a particular choice of the parameter $\varepsilon$, $\varepsilon=0$, 
 just the differential equation for Legendre polynomials. 

Both differential equations can easily be solved for the whole interval of the parameter $\varepsilon$,
$0\le \varepsilon< \frac{1}{2}$, due to the relations of the first
order differential 
equations~(\ref{Bnanzatzl0}),  once we find the coefficients
$a^{l}_{n}$ from the recursion 
relation~(\ref{aln}) and use them 
in the expression for $\mathcal{A}^{l_0}_n $,  $\mathcal{A}^{l_0}_n = (1+x)^{-\varepsilon}\, 
 \sum_{l=n}^{l_0} \, a^{l}_n\, P^{l}_n\,$~(\ref{Ananzatzl0}).
The normalisable solutions are expressible with a 
finite sum of the Legendre polynomials~(\ref{Ananzatzl0}, \ref{Bnanzatzl0}).

Let us now present solutions $\mathcal{A}^{l_0}_n$ and $\mathcal{B}^{l_0}_{n+1}$, 
needed to know the solution of~(\ref{mabpsil0}), for  particular choices of masses 
($(m\, \rho_{0})^2= 
l_{0}(l_{0} +1 - 2 \varepsilon)\,,$ $0 \le \varepsilon < \frac{1}{2}$)~(\ref{mass}), 
that is for a particular $l_{0}$.
Results are plotted with Mathematica. 
Since $\;\mathcal{A}^{l_0}_n$ are all singular 
at $x=-1$, but yet square integrable in the interval $0\le \varepsilon < \frac{1}{2}$, Mathematica 
makes the approximation with $0$ for very small values of $\;\mathcal{A}^{l_0}_n$ at $\varepsilon$
close to zero.

\begin{figure}
\centering
\includegraphics{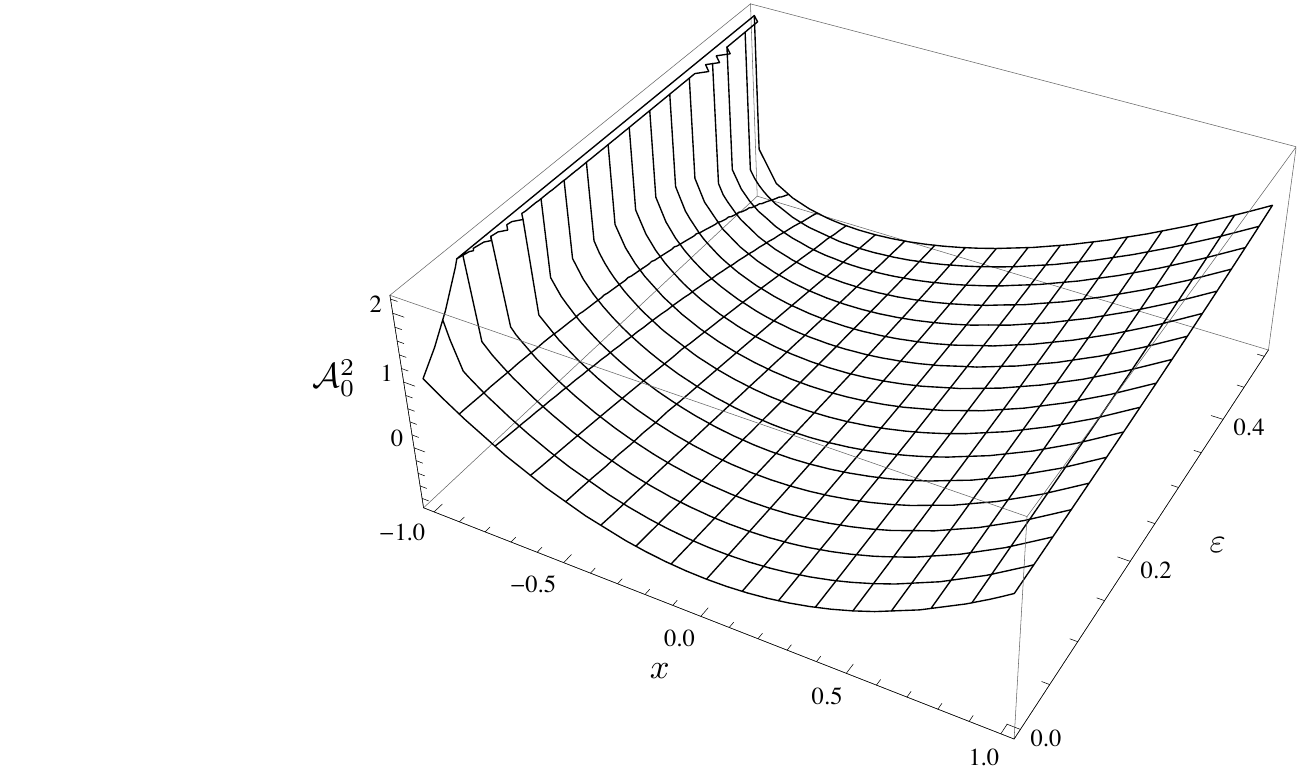}
\caption{The solution  $\mathcal{A}^{l_0}_0$\, 
of~(\ref{mainsecA}) for $l_0=2\,,\,n=0$, 
 is represented as a function of $x$, $-1\le x \le1$ for  
 $0\le \varepsilon < \frac{1}{2}$. We set $ a^{2}_0=1$. 
\label{l02n0Axeps}}
\end{figure}

In figure~\ref{l02n0Axeps}  the solution $\;\mathcal{A}^{l_0}_0$ of~(\ref{mainsecA}) for 
$l_{0}=2,\,n=0,\,$ is presented as a function of 
$x$, $-1\le x \le 1$,   and $\varepsilon$,  $\,0\le \varepsilon < \frac{1}{2}$. $\mathcal{A}^{l_0}_0$ 
changes from $\mathcal{A}^{2}_0 = P^{2}_0\,$ for $\varepsilon=0$ to the sum  of three 
Legendre polynomials, $P^{2}_{0}, P^{1}_{0}$ and $P^{0}_{0}$, 
 weighted by the coefficients $a^{2}_0,\,$ 
$a^{1}_0 = \frac{3\varepsilon}{2-\varepsilon}\,a^{2}_0\,,$ 
$\,a^{0}_0 =-\frac{2 \varepsilon\,(1-2 \varepsilon)}{(2-\varepsilon)(3-2 \varepsilon)}\, a^{2}_0$, 
respectively, and the function $(1+x)^{-\varepsilon}$ as written in~\Eref{Ananzatzl0sub}. 
In figure $a^{2}_0=1$ is taken. The mass is equal to $m\,\rho_{0}=\sqrt{2\,(3-2\varepsilon)}$.

\begin{figure}
\centering
\includegraphics{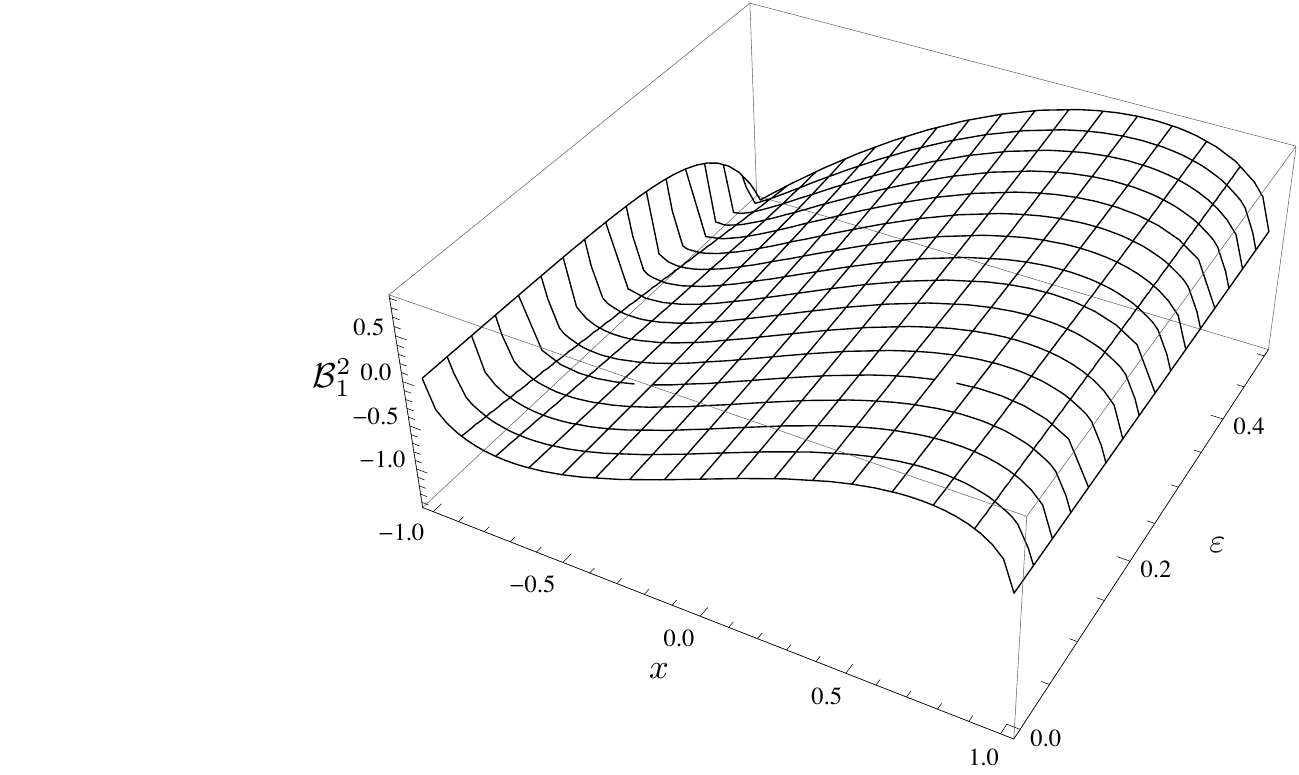}
\caption{The solution  $\mathcal{B}^{2}_{0+1}$ of~(\ref{Bnanzatzl0}), 
 which together with  $\mathcal{A}^{2}_0$ 
 from figure~\ref{l02n0Axeps} determines a spinor  state~(\ref{mabpsil0})  with 
the mass $(m\,\rho_0)^2 = 2(3-2 \varepsilon)$, is presented as a function of $x$, $-1\le x \le1$ for  
 $0\le \varepsilon < \frac{1}{2}$. 
\label{l02n0Bxeps}}
\end{figure}

In figure~\ref{l02n0Bxeps} the solution $\;\mathcal{B}^{l_0}_{0+1}$  for $l_{0}=2, n=0,$  
is presented. It determines,
together with $\mathcal{A}^{2}_0$ from figure~\ref{l02n0Axeps},  the spinor  state~(\ref{mabpsil0}) 
with the mass $m\,\rho_0 = \sqrt{2\,(3-2\varepsilon)}$. 
 The solution $\mathcal{B}^{2}_{0+1}$ changes from $\mathcal{B}^{2}_{0+1} = \frac{i \sqrt{6}}{5}\, 
\frac{1}{\sqrt{1-x^2}} \times  [ P^{3}_{0} -P^{1}_{0}]\,$ for $\varepsilon=0$ to the function
$\mathcal{B}^{2_0}_{0+1} = \frac{i (1+x)^{-\varepsilon}}{\sqrt{2(3-2\varepsilon)}\,\sqrt{1-x^2}} \times  
\, [ \frac{6}{5}\,(P^{3}_{0} -P^{1}_{0}) + \frac{2\varepsilon}{2-\varepsilon}\,(P^{2}_{0} -P^{0}_{0})]$ 
for $0\le \varepsilon<\frac{1}{2}$. 
While $\;\mathcal{A}^{2}_{0}$ 
is infinite at $x=-1$, but integrable, $\;\mathcal{B}^{l_0}_{0+1}$ is finite in the whole interval for 
any $l_{0}\,, n=0$, due to the fact that $((P^{l+1}_{0} -P^{l-1}_{0}))|_{x=-1} = 0$. This is no longer true if 
$n\ne 0$.

\begin{figure}
\centering
\includegraphics{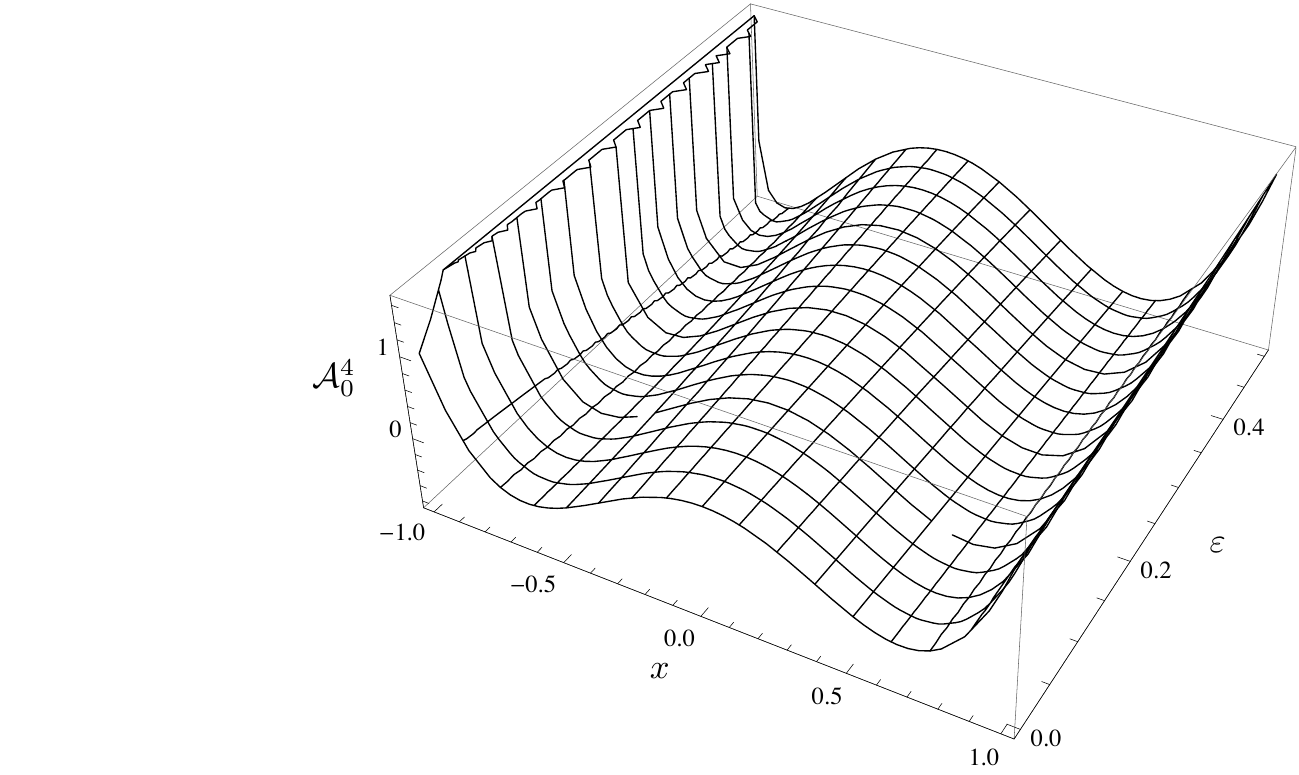}
\caption{ $\mathcal{A}^{4}_0$
, the solution of~(\ref{mainsecA}) for $l_0=4\,,n=0$, 
 is presented as a function of $x$, $-1\le x \le1$ for  
 $0\le \varepsilon < \frac{1}{2}$. We set $ a^{4}_0=1$. 
\label{l04n0Axeps}}
\end{figure}

\begin{figure}
\centering
\includegraphics{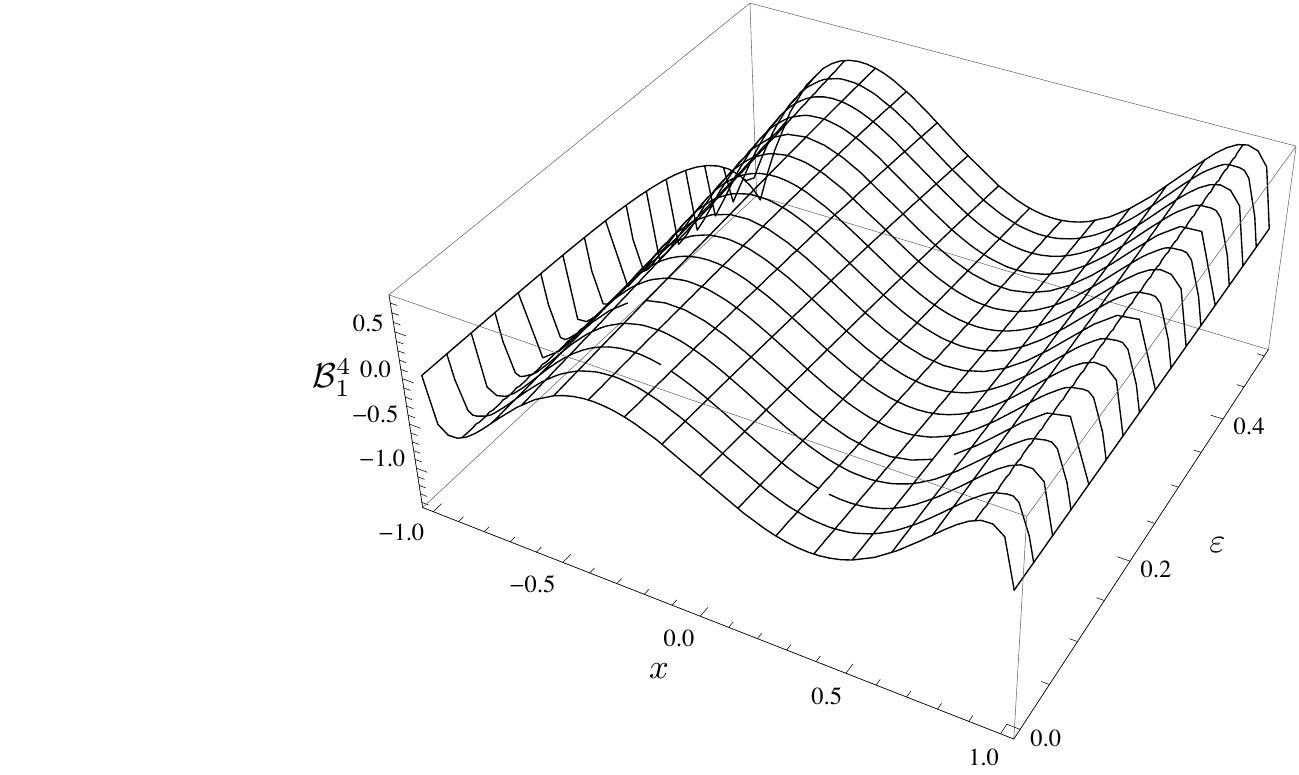}
\caption{ $\mathcal{B}^{4}_{0+1} $, which together with  $\mathcal{A}^{4}_0$ 
 from figure~\ref{l04n0Axeps} determines a spinor  state~(\ref{mabpsil0})  with 
the mass $(m \rho_0)^2 = 4(5-2 \varepsilon)$, is presented as a function of $x$, $-1\le x \le1$ for  
 $0\le \varepsilon < \frac{1}{2}$. 
\label{l04n0Bxeps}}
\end{figure}

\begin{figure}
  \centering
  \includegraphics{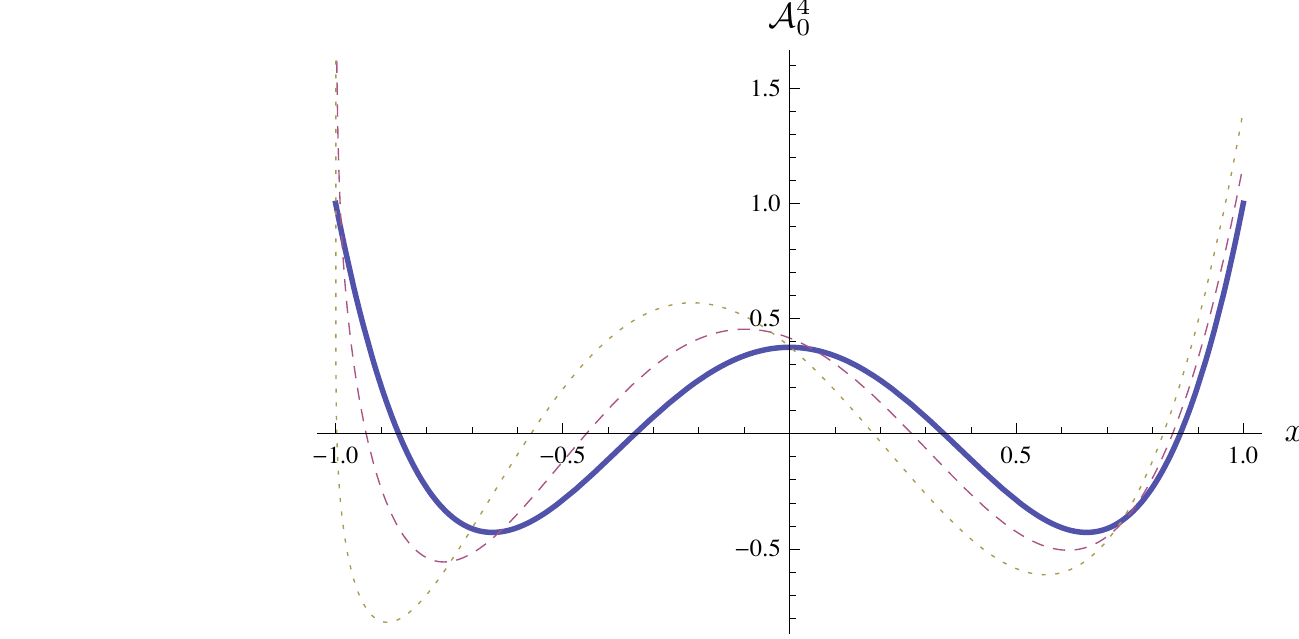}
  \caption{ $\mathcal{A}^{4}_0 $, 
  the solution of~(\ref{mainsecA}) for $l_0=4$ from figure~\ref{l04n0Axeps} for three values of 
  $\varepsilon=0, \,0.25,\, 0.49$ (thick, dashed and dotted, respectively) 
   is presented as a function of $x$, $-1\le x \le1$. 
   \label{l04n0Axepszeros}}
\end{figure}

\begin{figure}
   \centering
   \includegraphics{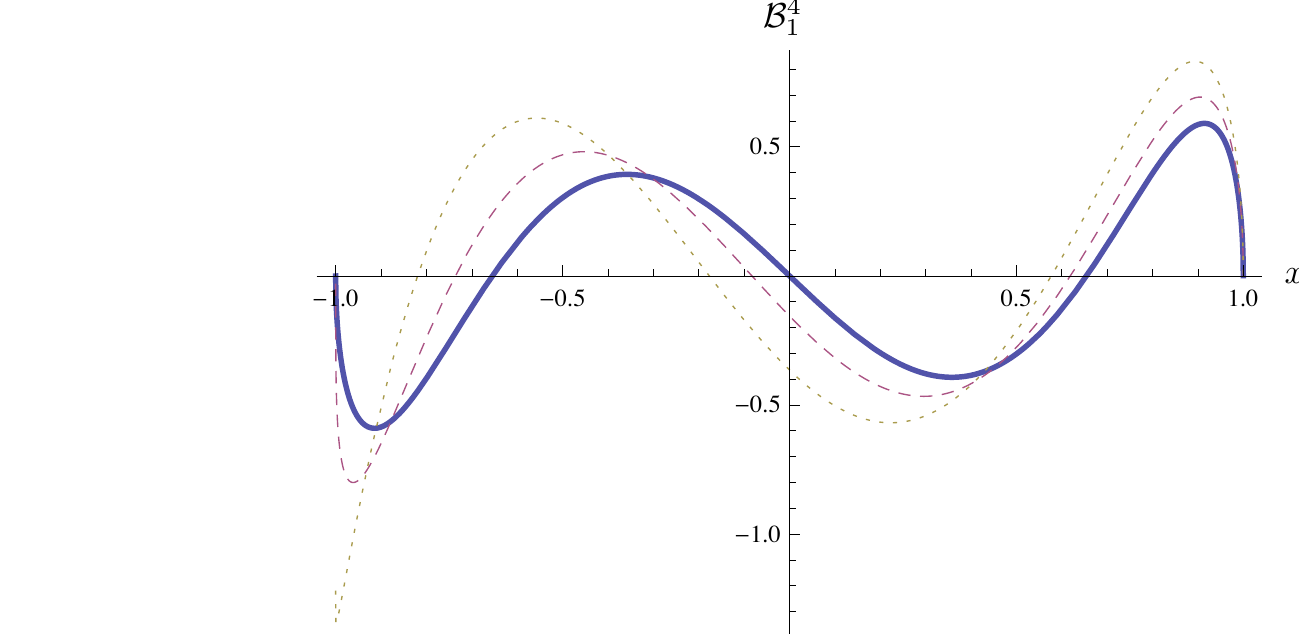}
   \caption{ $\mathcal{B}^{4}_0$ from figure~\ref{l04n0Bxeps} is presented as a function of 
    $x$, $-1\le x \le1$, for  three values of $\varepsilon =0, \,0.25,\,0.49$ 
    (thick, dashed and dotted, respectively). 
    \label{l04n0Bxepszeros}}
\end{figure}

In figures~\ref{l04n0Axeps} and~\ref{l04n0Bxeps}  the functions $\mathcal{A}^{4}_0$ and 
$\mathcal{B}^{4}_{0+1}$, which solve for $l_{0}= 4, n=0$, equation ~(\ref{maineps}), 
are presented.  
The zeros of both functions for three particular values of $\varepsilon$ can  be read  in 
figures~\ref{l04n0Axepszeros}, \ref{l04n0Bxepszeros}. 

Although the  functions  $\mathcal{A}^{l_0}_{n}$ and $\mathcal{B}^{l_0}_{n+1}$ are infinite at 
$x=-1$ for all $\varepsilon \ne 0$,  $\mathcal{A}^{l_0}_{n}$ for any $n \le l_{0}$ and 
 $\mathcal{B}^{l_0}_{n+1}$ for any $n\ne 0$ and $n \le l_{0}$, the solutions are all square 
integrable for $0\le \varepsilon < \frac{1}{2}$ in the interval $-1\le x \le1$ and correspondingly 
normalisable.


\begin{figure}
\centering
\includegraphics{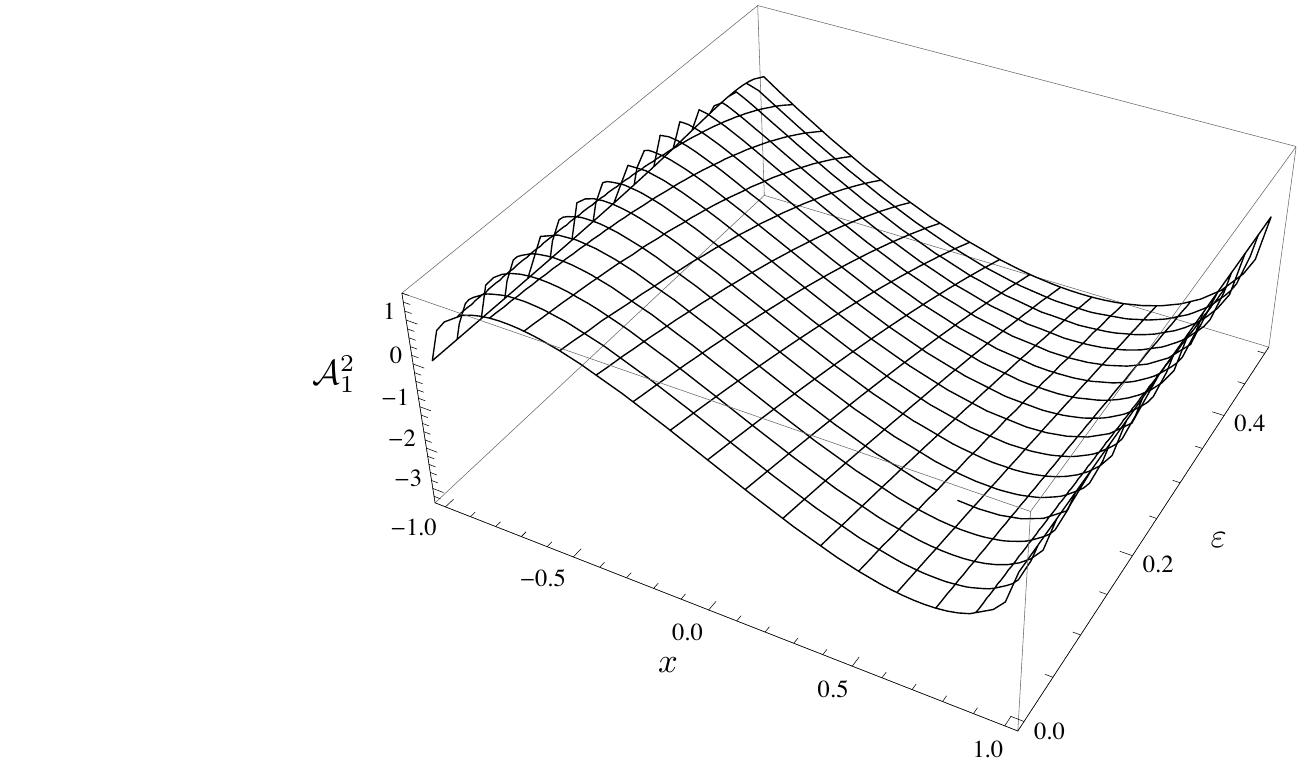}
\caption{ $\mathcal{A}^{2}_{1}$, the solution of~(\ref{mainsecA}) for $l_0=2$ and $n=1$, 
 is presented as a function of $x$, $\,-1\le x \le1$, for  
 $\,0\le \varepsilon < \frac{1}{2}$. We set $ a^{2}_0=1$. 
\label{l02-20Axeps}}
\end{figure}

\begin{figure}
\centering
\includegraphics{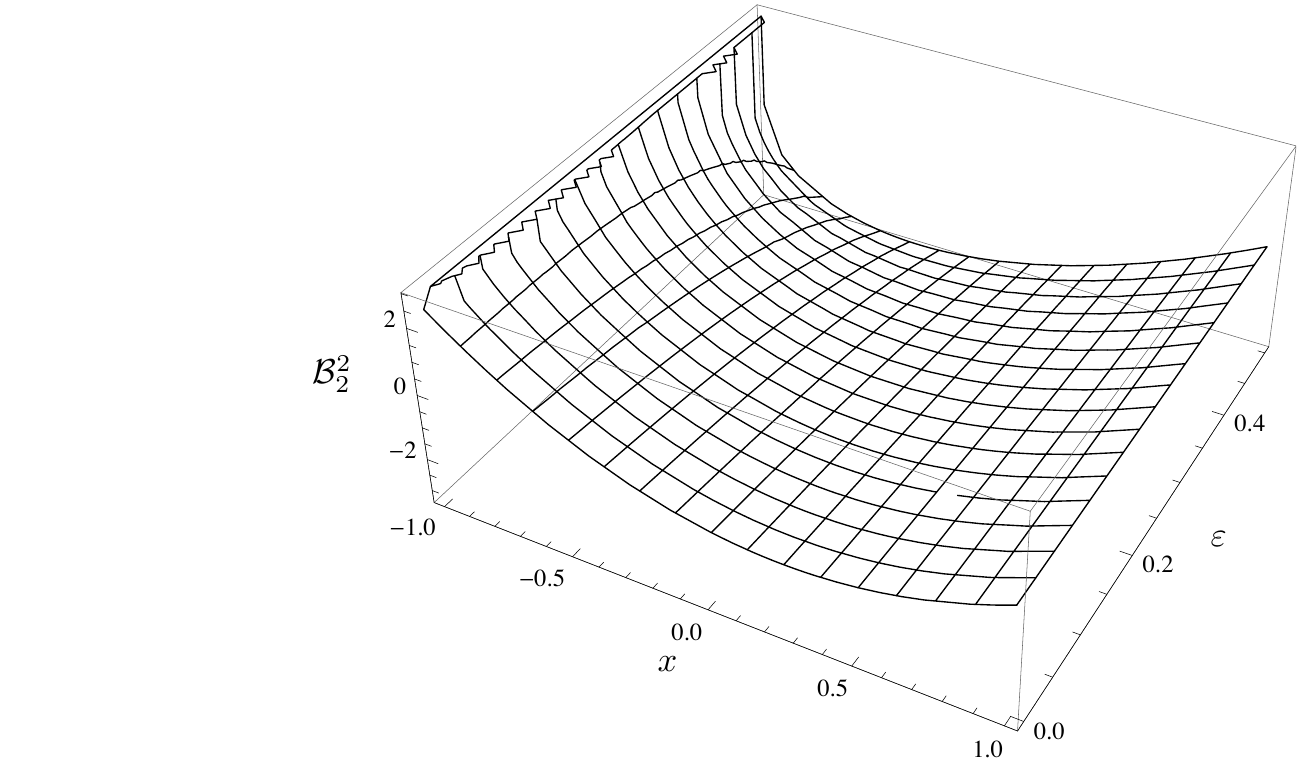}
\caption{$\mathcal{B}^{2}_{1+1}$, which together with  $\mathcal{A}^{2}_{1}$ 
 from figure~\ref{l02-20Axeps} determines a spinor  state~(\ref{mabpsil0})  with 
the mass $(m\,\rho_0)^2 = 2(3-2 \varepsilon)$, is represented as a function of $x$, $-1\le x \le1$ for  
 $0\le \varepsilon < \frac{1}{2}$. 
\label{l02-20Bxeps}}
\end{figure}

In figures~\ref{l02-20Axeps} and ~\ref{l02-20Bxeps}  we plot $\mathcal{A}^{2}_{1}$ and 
$\mathcal{B}^{2}_{1+1}\,$,  the solutions~(\ref{mabpsi}) of~(\ref{mainsecA}, \ref{mainsecB})  
for $l_{0}=2$  and $n=1$, as  functions of 
$x$, $-1\le x \le 1$  and $\varepsilon$,  $0\le \varepsilon < \frac{1}{2}$. 
The mass is  independent of the choice of $n$ and therefore equal to 
$m\,\rho_0= \sqrt{2(3-2 \varepsilon)}$. One finds $\mathcal{A}^{2}_{1}= (1+x)^{-\varepsilon}\,
a^{2}_{1} \, (P^{2}_{1} + 
\frac{9 \varepsilon}{2-\varepsilon} \,P^{1}_{1})$ and 
$\mathcal{B}^{2}_{1+1} = \frac{i (1+x)^{-\varepsilon}}{\sqrt{2(3-2\varepsilon)}\,\sqrt{1-x^2}} \times  
\, [ \frac{4}{5}\,P^{3}_{1} - (1 - \frac{3\varepsilon}{2-\varepsilon})\,P^{2}_{1} - 
9 \,(\frac{1}{5} + \frac{\varepsilon}{2-\varepsilon} ) \,P^{1}_{1}]$.


Let us conclude the paper by repeating briefly what we have succeeded to do. 
We prove  that the normalisable solutions of equations of motion~(\ref{weylE}) 
can be found for an interval $0\le \varepsilon < \frac{1}{2}$ of the parameter $\varepsilon =
\frac{1}{2}- F$, if a spinor lives on a manifold $M^{1+5}$, which breaks into $M^{1+3}\times$ 
an infinite disc, with the zweibein of~(\ref{f}) curving the disc into an almost $S^2$ and with the 
spin connection field~(\ref{omegas}) which allows on such a disc a massless spinor of only one 
handedness. The solutions were found by using the ansatz~(\ref{Ananzatzl0}), with the expansion over the 
associate Legendre polynomials, which  offer  normalisable solutions only for a finite sum, that is up to 
$l_0$, $\, l\le l_0$. For each $l_0=0,1,2,\dots$ and each  
chosen value of $0\le \varepsilon < \frac{1}{2}$ a discrete mass spectrum ($(m\,\rho_0)^2 = l_0\,(l_{0} +1 
-2 \varepsilon)$) follows, determined by $\rho_{0}$.  There is the gap between the massless 
and massive states as long as the zweibein and the spin connection remain unchanged. Correspondingly,  because 
of the gap, the massless state can not go continuously into the massive ones for any $0\le < \frac{1}{2}$. 
High enough temperatures in comparison 
with  $\frac{1}{\rho_{0}}$ can, however, since the space is non compact, cause the change  of the fields on the 
infinite disc and correspondingly also of the spectrum. But as long as the temperature is low, we take it 
zero in this paper, the massless spinor is (on the tree level) mass protected and also chirally coupled to the 
corresponding Kaluza-Klein gauge field.

The proof presented in this paper is a step to the proof that one can  escape from the "no-go theorem" of 
Witten~\cite{witten}, that is that one can guarantee the masslessness of spinors and their chiral coupling  to the 
Kaluza-Klein[like] gauge fields when breaking the symmetry from the $d$-dimensional one to 
$M^{(1+3)} \times M^{d-4}$ space.

The presented way of searching for solutions of the coupled equations~(\ref{maineps}) of the first order,  
leading to the  differential equations of the second order~(\ref{mainsecA}, \ref{mainsecB}), 
is by itself a nice presentation of how can one find normalisable solutions of quite complicated differential 
equations of the second order.

\ack  The authors acknowledge the financial support of the 
Slovenian Research Agency, Project P1-0188.

\appendix

\section{The technique for representing spinors~\cite{norma92939495,holgernorma20023}, 
taken from~\cite{NF}}
\label{technique}

The technique~\cite{norma92939495,holgernorma20023} can be used to construct a spinor basis for any dimension $d$
and any signature in an easy and transparent way. Equipped with the graphic presentation of basic states,  
the technique offers an elegant way to see all the quantum numbers of states with respect to the  
Lorentz groups, as well as transformation properties of the states under any Clifford algebra object.

The objects $\gamma^a$ 
have properties 
\begin{eqnarray}
\label{gammatildegamma}
&& \{ \gamma^a, \gamma^b\}_{+} = 2\eta^{ab}\,I, 
\end{eqnarray}
for any $d$, even or odd.  $I$ is the unit element in the Clifford algebra.

The Clifford algebra objects $S^{ab}$  
close the algebra of the Lorentz group 
\begin{eqnarray}
\label{sabtildesab}
S^{ab}: &=& (i/4) (\gamma^a \gamma^b - \gamma^b \gamma^a)\,, \nonumber\\
\{S^{ab},S^{cd}\}_{-} &=& i(\eta^{ad} S^{bc} + \eta^{bc} S^{ad} - \eta^{ac} S^{bd} - \eta^{bd} S^{ac})\,.
\end{eqnarray}

The ``Hermiticity'' property for $\gamma^a$'s  
\begin{eqnarray}
\gamma^{a\dagger} = \eta^{aa} \gamma^a\,,
\label{cliffher}
\end{eqnarray}
is assumed  in order that 
$\gamma^a$ 
are compatible with (\ref{gammatildegamma}) and formally unitary, 
i.e. $\gamma^{a \,\dagger} \,\gamma^a=I$. 

One finds from~(\ref{cliffher}) that $(S^{ab})^{\dagger} = \eta^{aa} \eta^{bb}S^{ab}$.

The Cartan subalgebra of the algebra 
is chosen as follows 
\begin{eqnarray}
S^{03}, S^{12}, S^{56}, \cdots, S^{d-1\; d}, \quad {\rm if } \quad d &=& 2n\ge 4,
\nonumber\\
S^{03}, S^{12}, \cdots, S^{d-2 \;d-1}, \quad {\rm if } \quad d &=& (2n +1) >4\,.
\label{choicecartan}
\end{eqnarray}

The choice for  the Cartan subalgebra in $d <4$ is straightforward.
It is  useful  to define one of the Casimirs of the Lorentz group -  
the  handedness $\Gamma$ ($\{\Gamma, S^{ab}\}_- =0$) in any $d$ 
\begin{eqnarray}
\Gamma^{(d)} :&=&(i)^{d/2}\; \;\;\;\;\;\prod_a \quad (\sqrt{\eta^{aa}} \gamma^a), \quad {\rm if } \quad d = 2n, 
\nonumber\\
\Gamma^{(d)} :&=& (i)^{(d-1)/2}\; \prod_a \quad (\sqrt{\eta^{aa}} \gamma^a), \quad {\rm if } \quad d = 2n +1\,.
\label{hand}
\end{eqnarray}
The product of $\gamma^a$'s in the ascending order with respect to 
the index $a$: $\gamma^0 \gamma^1\cdots \gamma^d$ is understood. 
It follows from~(\ref{cliffher})
for any choice of the signature $\eta^{aa}$ that
$\Gamma^{\dagger}= \Gamma,\;
\Gamma^2 = I.$
For $d$ even the handedness  anticommutes with the Clifford algebra objects 
$\gamma^a$ ($\{\gamma^a, \Gamma \}_+ = 0$), while for $d$ odd it commutes with  
$\gamma^a$ ($\{\gamma^a, \Gamma \}_- = 0$). 

To make the technique simple  the graphic presentation is introduced
\begin{eqnarray}
\stackrel{ab}{(k)}:&=& 
\frac{1}{2}(\gamma^a + \frac{\eta^{aa}}{ik} \gamma^b)\,,\quad \quad
\stackrel{ab}{[k]}:=
\frac{1}{2}(1+ \frac{i}{k} \gamma^a \gamma^b)\,,\nonumber\\
\stackrel{+}{\circ}:&=& \frac{1}{2} (1+\Gamma)\,,\quad \quad
\stackrel{-}{\bullet}:= \frac{1}{2}(1-\Gamma),
\label{signature}
\end{eqnarray}
where $k^2 = \eta^{aa} \eta^{bb}$.
One can easily check by taking into account the Clifford algebra relation 
(\ref{gammatildegamma}) and the
definition of $S^{ab}$ 
(\ref{sabtildesab})
that if one multiplies from the left hand side by $S^{ab}$ 
the Clifford algebra objects $\stackrel{ab}{(k)}$ and $\stackrel{ab}{[k]}$, it follows that
\begin{eqnarray}
        S^{ab}\, \stackrel{ab}{(k)}= \frac{1}{2}\,k\, \stackrel{ab}{(k)}\,,\quad \quad 
        S^{ab}\, \stackrel{ab}{[k]}= \frac{1}{2}\,k \,\stackrel{ab}{[k]}\,,
\label{grapheigen}
\end{eqnarray}
which means that we get the same objects back multiplied by the constant $\frac{1}{2}k$.
This also means that when 
$\stackrel{ab}{(k)}$ and $\stackrel{ab}{[k]}$ act from the left hand side on  a
vacuum state $|\psi_0\rangle$ the obtained states are the eigenvectors of $S^{ab}$.
One can further recognize 
that $\gamma^a$ transform  $\stackrel{ab}{(k)}$ into  $\stackrel{ab}{[-k]}$, never to $\stackrel{ab}{[k]}$:  
\begin{eqnarray} \fl
\gamma^a \stackrel{ab}{(k)}&=& \eta^{aa}\stackrel{ab}{[-k]},\; 
\gamma^b \stackrel{ab}{(k)}= -ik \stackrel{ab}{[-k]}, \; 
\gamma^a \stackrel{ab}{[k]}= \stackrel{ab}{(-k)},\; 
\gamma^b \stackrel{ab}{[k]}= -ik \eta^{aa} \stackrel{ab}{(-k)}\,.
\label{snmb:gammatildegamma}
\end{eqnarray}
From~(\ref{snmb:gammatildegamma}) it follows
\begin{eqnarray}
\label{stildestrans}
S^{ac}\stackrel{ab}{(k)}\stackrel{cd}{(k)} &=& -\frac{i}{2} \eta^{aa} \eta^{cc} 
\stackrel{ab}{[-k]}\stackrel{cd}{[-k]}\,,
\nonumber\\
S^{ac}\stackrel{ab}{[k]}\stackrel{cd}{[k]} &=& \frac{i}{2}  
\stackrel{ab}{(-k)}\stackrel{cd}{(-k)}\,,
\nonumber\\
S^{ac}\stackrel{ab}{(k)}\stackrel{cd}{[k]}  &=& -\frac{i}{2} \eta^{aa}  
\stackrel{ab}{[-k]}\stackrel{cd}{(-k)}\,,
\nonumber\\
S^{ac}\stackrel{ab}{[k]}\stackrel{cd}{(k)} &=& \frac{i}{2} \eta^{cc}  
\stackrel{ab}{(-k)}\stackrel{cd}{[-k]}\,. 
\end{eqnarray}

Let us deduce some useful relations
\begin{eqnarray}
\fl \stackrel{ab}{(k)}\stackrel{ab}{(k)}& =& 0\,, \quad \quad \stackrel{ab}{(k)}\stackrel{ab}{(-k)}
= \eta^{aa}  \stackrel{ab}{[k]}\,, \quad \stackrel{ab}{(-k)}\stackrel{ab}{(k)}=
\eta^{aa}   \stackrel{ab}{[-k]}\,,\quad
\stackrel{ab}{(-k)} \stackrel{ab}{(-k)} = 0\,, \nonumber\\
\fl \stackrel{ab}{[k]}\stackrel{ab}{[k]}& =& \stackrel{ab}{[k]}\,, \quad \quad
\stackrel{ab}{[k]}\stackrel{ab}{[-k]}= 0\,, \;\;\quad \quad  \quad \stackrel{ab}{[-k]}\stackrel{ab}{[k]}=0\,,
 \;\;\quad \quad \quad \quad \stackrel{ab}{[-k]}\stackrel{ab}{[-k]} = \stackrel{ab}{[-k]}\,,
 \nonumber\\
\fl \stackrel{ab}{(k)}\stackrel{ab}{[k]}& =& 0\,,\quad \quad \quad \stackrel{ab}{[k]}\stackrel{ab}{(k)}
=  \stackrel{ab}{(k)}\,, \quad \quad \quad \stackrel{ab}{(-k)}\stackrel{ab}{[k]}=
 \stackrel{ab}{(-k)}\,,\quad \quad \quad 
\stackrel{ab}{(-k)}\stackrel{ab}{[-k]} = 0\,,
\nonumber\\
\fl \stackrel{ab}{(k)}\stackrel{ab}{[-k]}& =&  \stackrel{ab}{(k)}\,,
\quad \quad \stackrel{ab}{[k]}\stackrel{ab}{(-k)} =0,  \quad \quad 
\quad \stackrel{ab}{[-k]}\stackrel{ab}{(k)}= 0\,, \quad \quad \quad \quad
\stackrel{ab}{[-k]}\stackrel{ab}{(-k)} = \stackrel{ab}{(-k)}.
\label{graphbinoms}
\end{eqnarray}
We recognize in  the first equation of the first row and the first equation of the second row
the demonstration of the nilpotent and the projector character of the Clifford algebra objects 
$\stackrel{ab}{(k)}$ and $\stackrel{ab}{[k]}$, respectively. 
Recognizing that
\begin{eqnarray}
\stackrel{ab}{(k)}^{\dagger}=\eta^{aa}\stackrel{ab}{(-k)}\,,\quad
\stackrel{ab}{[k]}^{\dagger}= \stackrel{ab}{[k]}\,,
\label{graphherstr}
\end{eqnarray}
 a vacuum state $|\psi_0>$ can be defined so that it follows
\begin{eqnarray}
< \;\stackrel{ab}{(k)}^{\dagger}
 \stackrel{ab}{(k)}\; > = 1\,,\quad \quad < \;\stackrel{ab}{[k]}^{\dagger}
 \stackrel{ab}{[k]}\; > = 1\,.
\label{graphherscal}
\end{eqnarray}

Taking into account the above equations it is easy to find a Weyl spinor irreducible representation
for $d$-dimensional space, with $d$ even or odd.

For $d$ even we simply make a starting state as a product of $d/2$, let us say, only nilpotents 
$\stackrel{ab}{(k)}$, one for each $S^{ab}$ of the Cartan subalgebra  elements (\ref{choicecartan}),  
applying it on an (unimportant) vacuum state. 
For $d$ odd the basic states are products
of $(d-1)/2$ nilpotents and a factor $(1\pm \Gamma)$.  
Then the generators $S^{ab}$, which do not belong 
to the Cartan subalgebra, being applied on the starting state from the left, 
 generate all the members of one
Weyl spinor.  
\begin{eqnarray}
\stackrel{0d}{(k_{0d})} \stackrel{12}{(k_{12})} \stackrel{35}{(k_{35})}\cdots \stackrel{d-1\;d-2}{(k_{d-1\;d-2})}
\psi_0 \nonumber\\
\stackrel{0d}{[-k_{0d}]} \stackrel{12}{[-k_{12}]} \stackrel{35}{(k_{35})}\cdots \stackrel{d-1\;d-2}{(k_{d-1\;d-2})}
\psi_0 \nonumber\\
\stackrel{0d}{[-k_{0d}]} \stackrel{12}{(k_{12})} \stackrel{35}{[-k_{35}]}\cdots \stackrel{d-1\;d-2}{(k_{d-1\;d-2})}
\psi_0 \nonumber\\
\vdots \nonumber\\
\stackrel{0d}{[-k_{0d}]} \stackrel{12}{(k_{12})} \stackrel{35}{(k_{35})}\cdots \stackrel{d-1\;d-2}{[-k_{d-1\;d-2}]}
\psi_0 \nonumber\\
\stackrel{od}{(k_{0d})} \stackrel{12}{[-k_{12}]} \stackrel{35}{[-k_{35}]}\cdots \stackrel{d-1\;d-2}{(k_{d-1\;d-2})}
\psi_0 \nonumber\\
\vdots 
\label{graphicd}
\end{eqnarray}
All the states have the handedness $\Gamma $, since $\{ \Gamma, S^{ab}\}-{+} = 0$. 
States, belonging to one multiplet  with respect to the group $SO(q,d-q)$, that is to one
irreducible representation of spinors (one Weyl spinor), can have any phase. We made a choice
of the simplest one, taking all  phases equal to one.

The above graphic representation demonstrates that for $d$ even 
all the states of one irreducible Weyl representation of a definite handedness follow from a starting state, 
which is, for example, a product of nilpotents $\stackrel{ab}{(k_{ab})}$, by transforming all possible pairs
of $\stackrel{ab}{(k_{ab})} \stackrel{mn}{(k_{mn})}$ into $\stackrel{ab}{[-k_{ab}]} \stackrel{mn}{[-k_{mn}]}$.
There are $S^{am}, S^{an}, S^{bm}, S^{bn}$, which do this.
The procedure gives $2^{(d/2-1)}$ states. A Clifford algebra object $\gamma^a$ being applied from the left hand side,
transforms  a 
Weyl spinor of one handedness into a Weyl spinor of the opposite handedness. Both Weyl spinors form a Dirac 
spinor.

For $d$ odd a Weyl spinor has besides a product of $(d-1)/2$ nilpotents or projectors also either the
factor $\stackrel{+}{\circ}:= \frac{1}{2} (1+\Gamma)$ or the factor
$\stackrel{-}{\bullet}:= \frac{1}{2}(1-\Gamma)$.  
As in the case of $d$ even, all the states of one irreducible 
Weyl representation of a definite handedness follow from a starting state, 
which is, for example, a product of $(1 + \Gamma)$ and $(d-1)/2$ nilpotents $\stackrel{ab}{(k_{ab})}$, by 
transforming all possible pairs
of $\stackrel{ab}{(k_{ab})} \stackrel{mn}{(k_{mn})}$ into $\stackrel{ab}{[-k_{ab}]} \stackrel{mn}{[-k_{mn}]}$.
But $\gamma^a$'s, being applied from the left hand side, do not change the handedness of the Weyl spinor, 
since $\{ \Gamma,
\gamma^a \}_- =0$ for $d$ odd. 
A Dirac and a Weyl spinor are for $d$ odd identical and a ''family'' 
has accordingly $2^{(d-1)/2}$ members of basic states of a definite handedness.

We shall speak about left handedness when $\Gamma= -1$ and about right
handedness when $\Gamma =1$ for either $d$ even or odd.

Making a choice of the Cartan subalgebra set of the algebra $S^{ab}$ 
\begin{eqnarray}
S^{03}, S^{12}, S^{56}\,, 
\label{cartan}
\end{eqnarray}
a left handed ($\Gamma^{(1,5)} =-1$) eigen state of all the members of the 
Cartan  subalgebra 
can be written as 
\begin{eqnarray}
&& \stackrel{03}{(+i)}\stackrel{12}{(+)} \stackrel{56}{(+)}|\psi \rangle = 
\frac{1}{2^3} 
(\gamma^0 -\gamma^3)(\gamma^1 +i \gamma^2)| (\gamma^5 + i\gamma^6)
|\psi \rangle \,.
\label{start}
\end{eqnarray}
This state is an eigen state of all $S^{ab}$ 
which are members of the Cartan 
subalgebra (\ref{cartan}).

Below some useful relations~\cite{norma92939495} are presented 
\begin{eqnarray}
\label{plusminus}
N^{\pm}_{+}         &=& N^{1}_{+} \pm i \,N^{2}_{+} = 
 - \stackrel{03}{(\mp i)} \stackrel{12}{(\pm )}\,, \quad N^{\pm}_{-}= N^{1}_{-} \pm i\,N^{2}_{-} = 
  \stackrel{03}{(\pm i)} \stackrel{12}{(\pm )}\,,\nonumber\\
\end{eqnarray}

\section{Useful relations among Legendre polynomials}
\label{legendre}

We present in this appendix some useful relations, some of them are well known~\cite{WG}
\begin{eqnarray}
\quad x P^{l}_n &=&\frac{1}{2l+1} \left[ (l+n) P^{l-1}_n + (l-n+1) P^{l+1}_n\right]\,,
\label{eq:xPln}
\end{eqnarray}
\begin{eqnarray}
(2 l +1) \sqrt{1-x^2} P^l_{n-2} = P^{l-1}_{n-1} - P^{l+1}_{n-1}\,,\nonumber\\
\fl
(2 l +1) \sqrt{1-x^2} P^l_{n} = (l-n+2)(l-n+1) P^{l+1}_{n-1} \nonumber\\
 -(l+n)(l+n-1) P^{l-1}_{n-1}\,,
\end{eqnarray}
\begin{eqnarray}
(1-x^2) \frac{d}{dx} P^l_n &=&
\frac{1}{2l+1}\left[  (l+n)(l+1) P^{l-1}_n - l(l-n+1) P^{l+1}_n\right]\,.
\label{eq:1mx2dxPln}
\end{eqnarray}
Others follow with some effort from the above ones.
\begin{eqnarray}
\fl 
(1-x^2) P^l_n = \frac{1}{2l+1} \biggl\{
   \left[  2l+1 - \frac{(l+n)(l-n)}{2l-1} -
     \frac{(l-n+1)(l+n+1)}{2l+3} \right] P^l_n \nonumber\\
 {}-\frac{(l+n)(l+n-1)}{2l-1} P^{l-2}_n 
  - \frac{(l-n+1)(l-n+2)}{2l+3} P^{l+2}_n 
\biggr\}\,, \label{eq:1mx2Pln}
\end{eqnarray}
\begin{eqnarray}
&&\frac{1}{\sqrt{1-x^2}}P^l_n = -\frac{1}{2n} \left( (l-n+2)(l-n+1)
  P^{l+1}_{n-1}+ P^{l+1}_{n+1}\right)\,,\label{eq:1sqrtPln1} \\
&&\frac{1}{\sqrt{1-x^2}}P^l_n = -\frac{1}{2n} \left( (l+n)(l+n-1)
  P^{l-1}_{n-1}+ P^{l-1}_{n+1}\right)\,.\label{eq:1sqrtPln2}
\end{eqnarray}
Any function $f$, continuous on the open interval $(-1,1)$ and square integrable in the closed  
 interval ($\int^{1}_{-1}\, dx \,|f|^{2}$), can be expanded in terms of associate 
 Legendre functions  $P^{l}_n$:
\begin{eqnarray}
\label{anyf}
f(x) &=& \sum_{l \ge n}\, c_{l}^{n} \,P^{l}_n \,(x)\,,\nonumber\\
c_{l}^{n}&=& \frac{2 l+1}{2}\,\frac{(l-n)!}{(l+n)!} \,\int_{-1}^{1} \,f(x)\,P^{l}_n\,(x)\,dx\,.
\end{eqnarray}
\section{Recursive relations for the ansatzes 
  $\mathcal{A}_{n}=\sum_{l\ge n}\, \alpha^l_n P^l_{n}$ and  
$\mathcal{B}_{n+1}=\sum_{l \ge n+1} \beta^{l}_{n+1} P^{l}_{n+1}$}
\label{usefulequations} 

Using relations from~\ref{legendre} it is not too difficult to find the recurrence relations for 
$\alpha^{l}_{n}$ when the ansatz $\mathcal{A}_{n}=\sum_{l \ge n}\, \alpha^{l}_n P^{l}_{n}$ is used in equation
\begin{eqnarray}
\fl
(1 - x^2) \frac{d^2}{dx^2}\mathcal{A}_n -2 x \frac{d}{dx}\mathcal{A}_n +(m\rho_0)^2 \mathcal{A}_n\nonumber\\
-\left\{\frac{n^2}{1-x^2} +\frac{1}{1+x} \varepsilon (1 + x +2 n)  + 
  \varepsilon^2 \frac{1-x}{1+x} \right\}\mathcal{A}_n =0\,.
  \end{eqnarray}
One finds
\begin{eqnarray}
\label{alpha}
\fl
\alpha^{l+1}_{n}  \frac{l+n+1}{2l+3}\left[
  (m\rho_0)^2-(l+1)(l+2)-\varepsilon(1-\varepsilon)\right]=\nonumber\\
-\alpha^l_n\left[
  (m\rho_0)^2-l(l+1)-\varepsilon(1+\varepsilon+2n)\right]\nonumber\\
-\alpha^{l-1}_n \frac{l-n}{2l-1} \left[  (m\rho_0)^2-(l-1)l-\varepsilon(1-\varepsilon)\right]\,.
\end{eqnarray}
Similarly one finds recurrence relations for $\beta^l_{n+1}$  when using  the ansatz 
$\mathcal{B}_{n+1}=\sum_{l\ge n+1} \beta^l_{n+1} P^l_{n+1}$ in equation 
\begin{eqnarray}
\fl
(1 - x^2) \frac{d^2}{dx^2}\mathcal{B}_{n+1} - 2 x
\frac{d}{dx}\mathcal{B}_{n+1} +(m\rho_0)^2 \mathcal{B}_{n+1}\nonumber\\
 +\left\{-\frac{ (n+1)^2}{1 - x^2}  
+\frac{2n(1-\varepsilon)}{1+x} 
+\varepsilon(1-\varepsilon)\frac{1-x}{1+x} 
\right\} \mathcal{B}_{n+1}=0\,.
\end{eqnarray}
One obtains
\begin{eqnarray}
\label{beta}
\fl
\beta^{l+1}_{n+1} \frac{l+n+2}{2l+3} \left[
  (m\rho_0)^2-(l+1)(l+2)-\varepsilon(1-\varepsilon)\right]=\nonumber\\
 -\beta^{l}_{n+1} \left[  (m\rho_0)^2-l(l+1)+(1-\varepsilon)
   (\varepsilon+2n)\right]\nonumber\\
-\beta^{l-1}_{n+1} \frac{l-n-1}{2l-1} \left[  (m\rho_0)^2-(l-1)l-\varepsilon(1-\varepsilon)\right]\,.
\end{eqnarray}

The two ansatzes are not really useful, since it is very difficult to evaluate for which 
values of $(m\rho_0)^2$ are the solutions square integrable. 
The ansatzes from~(\ref{Ananzatzl0}, \ref{Bnanzatzl0}) are much more appropriate 
offering us the solutions in quite an elegant way.


\begin{thebibliography}{99}
%
 %
 \bibitem{NHD} Lukman D, Manko\v c Bor\v stnik N S and  Nielsen H B
   2011 \NJP {\bf 13} (2011) 103027 
\bibitem{kk} Kaluza T 1921 {\it  Sitzungsber.Preuss.Akad.Wiss.Berlin, Math.Phys.} {\bf K1}  966,
Klein ) 1926 {\it Z.Phys.} {\bf 37}  895
%
\bibitem{geogla} Georgi H and Glashow S L 1974  \PRL {\bf 32}  438
\bibitem{chofre} Cho Y M  1975 {\it J. Math. Phys.} {\bf 16}  2029,
Cho Y M and Freund P G O 1975 \PR D {\bf 12} 1711 
\bibitem{zee} Zee A 1981  {\it Proc. 1st Kyoto Summer Institute on 
Grand Unified Theories and Related Topics (Kyoto)} ed M Konuma and T Kaskawa
(Singapore: World Scientific)  
%
\bibitem{salstr}  Salam A and  Strathdee J 1982 \APNY  {\bf 141} 316 
%
\bibitem{mec} Mecklenburg W 1984 {\it Fortschr. Phys.} {\bf 32} 207
%
\bibitem{dufnilspop} Duff M, Nilsson, B and Pope C 1984  {\it Phys. Rep.} C  {\bf C 130} 1,
Duff M, Nilsson B, Pope C and Warner N 1984 \PL B {\bf 149} 60
%
%
\bibitem{daess} Randjbar-Daemi S, Salam A and Strathdee J 1984 \NP B {\bf 242} 447
%
%
\bibitem{wet} Wetterich C 1985  {\it The 2nd Jerusalem Winter School on Theoretical Physics}
CERN-TH4190/85 and references therein, Wetterich C 1985 \NP B {\bf 253} 261, 
Wetterich C 1984 \NP B {\bf 234} 413
%
\bibitem{zelenaknjiga} The authors of the works presented in 1983 {\it An Introduction to Kaluza-Klein 
Theories} ed H C Lee (Singapore: World Scientific)
%
\bibitem{horpal} Horvath Z, Palla L, Crammer E and Scherk J 1977 \NP B
  {\bf 127} 57 
%
\bibitem{witten} Witten E 1981 
\NP B {\bf 186} 412; 
Witten E 1883 {\it Princeton Technical Rep. PRINT -83-1056, October 1983}
%
%
%
%

%
%
\bibitem{norma92939495} Manko\v c Bor\v stnik N S 1992  
\PL B {\bf 292} 25,
%
Manko\v c Bor\v stnik N S 1993 \JMP {\bf 34} 3731,
%
Manko\v c Bor\v stnik N S2001 {\it Int. J. Theor. Phys.} {\bf 40} 315,
%
Manko\v c Bor\v stnik N S 1995 {\it Modern Phys. Lett.} A  {\bf 10} 587, 
%
Bor\v stnik A and Manko\v c Bor\v stnik N S 2004 ({\it Preprint}  
hep-ph/0401043), Bor\v stnik A and Manko\v c Bor\v stnik N S 2004
({\it Preprint}  hep-ph/0401055) pp 27--51,
Bor\v stnik A and Manko\v c Bor\v stnik N S 2002 ({\it Preprint}   hep-ph/0301029)
%
Bor\v stnik A and Manko\v c Bor\v stnik N S 2006\PR D { \bf  74}
073013 ({\it Preprint} hep-ph/0512062), 
%
Bregar G, Breskvar M, Lukman D and Manko\v c Bor\v stnik N S 2007
({\it Preprint}   hep-ph/0711.4681) pp 53--70,
		    %
Bregar G, Breskvar M, Lukman D and Manko\v c Bor\v stnik N S 2008
 \NJP {\bf 10} 093002,  
Breskvar M, Lukman D and Manko\v c Bor\v stnik N S 2006 ({\it
  Preprint} hep-ph/0606159, 
Bregar G, Breskvar M, Lukman D and Manko\v c Bor\v stnik N S 2007
({\it Preprint} hep-ph/07082846), 
Breskvar M, Lukman D and Manko\v c Bor\v stnik N S 2006 ({\it
  Preprint} hep-ph/0612250) pp25--50
%
%
%
\bibitem{holgernorma20023} Manko\v c Bor\v stnik N S and Nielsen H B 2002
\JMP   {\bf 43} 5782, Manko\v c Bor\v stnik N S and Nielsen H B 2001,
Manko\v c Bor\v stnik N S and Nielsen H B 2001 ({\it Preprint} hep-th/0111257),               
%
Manko\v c Bor\v stnik N S and Nielsen H B 2003 \JMP  {\bf 44} 4817, 
Manko\v c Bor\v stnik N S and Nielsen H B 2003 ({\it Preprint} hep-th/0303224)
%
 %
\bibitem{hnkk06} Manko\v c Bor\v stnik N S and Nielsen H B 2006 
\PL B {\bf  633}  771--5, Manko\v c Bor\v stnik N S and Nielsen H B
2003 ({\it Preprint}  hep-th/0311037), Manko\v c Bor\v stnik N S and
Nielsen H B 2005 ({\it Preprint} hep-th/0509101),
%
Manko\v c Bor\v stnik N S and Nielsen H B 2007 \PL B {\bf 644} 198--202, 
Manko\v c Bor\v stnik N S and Nielsen H B 2006 ({\it Preprint} hep-th/0608006), 
%
Manko\v c Bor\v stnik N S and Nielsen H B 2008 \PL B {\bf 663} 265--96,
%
 Manko\v c Bor\v stnik N S, Nielsen H B and Lukman D 2004 
({\it Preprint} hep-ph/0412208)
%
%
\bibitem{WG} Wang Z X and Guo D R 1989 {\it Special Functions} (Singapore: Worls Scientific) 
 pp255--8
%
\bibitem{CS} Tikhonov A N and Samarskii  A A  1963 {\it Equations of Mathematical Physics},
({\it International Series of Monographs on Pure and Applied
  Mathematics} vol 39) (Oxford etc.: Pergamon Press)
%
\bibitem{NF}  Manko\v c Bor\v stnik N S 2010 ({\it Preprint} arxiv:1011.5765),
Manko\v c Bor\v stnik N S 2010 ({\it Preprint} arXiv:1012.0224) pp105--30 
%
\end{thebibliography}
\end{document}